\documentclass[3p,times]{elsarticle}

\usepackage{graphicx}
\usepackage{caption}
\usepackage{subcaption}
\usepackage{comment}
\usepackage{amsmath}
\usepackage[version=4]{mhchem}
\usepackage{float}
\usepackage{framed}
\usepackage[normalem]{ulem}
\usepackage{color}
\usepackage{mdframed}
\usepackage{longtable}
\usepackage{algorithm}
\usepackage{epstopdf}
\usepackage[noend]{algpseudocode}

\newcommand{\ul}[1]{\underline{#1}}

\usepackage{amssymb}

\usepackage[figuresright]{rotating}

\begin{document}

\begin{frontmatter}

\title{PyCDT:  A Python toolkit for modeling point defects in semiconductors and insulators} 

\author[1]{Danny Broberg\corref{author}\fnref{equal}}
\cortext[author]{Corresponding authors: D.Broberg (dbroberg@berkeley.edu), 
B. Medasani (bharat.medasani@pnnl.gov), N. Zimmermann (nerz@lbl.gov) 
and G. Hautier (geoffroy.hautier@uclouvain.be).}
\fntext[equal]{These authors contributed equally to this work.}
\author[2,3]{Bharat Medasani\corref{author}\fnref{equal}}
\author[2]{Nils Zimmermann\corref{author}\fnref{equal}}
\author[2]{Andrew Canning}
\author[2]{Maciej Haranczyk}
\author[1]{Mark Asta}
\author[4]{Geoffroy Hautier\corref{author}}

\address[1]{Department of Materials Science and Engineering, University of California, Berkeley, CA 94720, USA}
\address[2]{Computational Research Division, Lawrence Berkeley National Laboratory, Berkeley, CA 94720, USA}
\address[3]{Physical and Computational Sciences Directorate, Pacific Northwest National Laboratory, Richland, WA 99354, USA }
\address[4]{Institute of Condensed Matter and Nanosciences (IMCN), Universit\'{e} catholique de Louvain, 1348 Louvain-la-Neuve, Belgium}

\begin{abstract}

Point defects have a strong impact on the performance of
semiconductor and insulator materials 
used in technological applications,
spanning microelectronics to energy conversion and storage.
The nature of the dominant defect types,
how they vary with processing conditions,
and their impact on materials properties are central 
aspects that determine the performance of a material in a
certain application.
This information is, however, difficult
to access directly from experimental measurements.
Consequently, computational methods,
based on electronic density functional theory (DFT),
have found widespread use in the calculation of point-defect properties.
Here we have developed 
the Python Charged Defect Toolkit (PyCDT) to expedite the setup
and post-processing of defect calculations with widely used DFT software.
PyCDT has a 
user-friendly command-line interface and provides a direct interface with the Materials Project database.  This allows for setting up many
charged defect calculations for any material of interest, as well as 
post-processing and applying state-of-the-art
electrostatic correction terms. 
Our paper serves as a documentation for PyCDT,
and demonstrates its use in an application to the well-studied GaAs compound semiconductor.
We anticipate that the PyCDT code will be useful as a framework
for undertaking readily reproducible calculations
of charged point-defect properties,
and that it will provide a foundation for automated,
high-throughput calculations.

\begin{keyword}
Point Defects, Charged Defects, Semiconductors, Insulators, Density Functional Theory, Python

\end{keyword}

\end{abstract}

\end{frontmatter}

\vspace{0.8cm}
\noindent
{\bf PROGRAM SUMMARY}

\begin{small}
\noindent
{\em Program title:}    PyCDT                                   \\
{\em Licensing Provisions:} MIT License. \\
{\em Program obtainable from: https://bitbucket.org/mbkumar/pycdt}\\
{\em Distribution format:} Git repository \\
{\em Programming language:}    Python                               \\
{\em Computer:}    Any computer with a Python interpreter;                                           \\
{\em RAM:} Problem dependent         \\
{\em External routines/libraries:}
NumPy~\cite{numpy}, matplotlib~\cite{matplotlib}, and Pymatgen~\cite{ong2013python}, \\
{\em Nature of problem:} Computing the formation energies and stable point defects with finite size supercell error corrections for charged defects in semiconductors and insulators \\
{\em Solution method:}  Automated setup, and parsing of defect calculations, combined with local use of finite size supercell corrections. All combined into a code with a standard user-friendly command line interface that leverage a core set of tools with a wide range of applicability \\
{\em Running time:}
Problem dependent\\
{\em Additional comments:} This article describes version 1.0.0. \\
   \\
\end{small}

\section{Introduction}

Point defects in semiconductors and insulators govern a range of mechanical, 
transport, electronic, and optoelectronic
properties~\cite{callister2007, queisser2013,mccluskey2012,rodnyi1997,seebauer2006}.
Due to the fact that the properties of these defects are difficult to characterize fully from experiment~\cite{seebauer2006, freysoldt_rmp}, computational tools have been widely applied.
Many applications
such as 
lanthanide-doped scintillator materials~\cite{dorenbos2005,chaudhry2014},
transparent conducting oxide materials~\cite{zunger_TCO, scanlon_TCO, varley_TCO},
photovoltaic materials~\cite{zakutayev_solar, walsh_solar},
and new thermoelectric materials~\cite{zhu_te, pomrehn_te} have benefited from leveraging theory 
for calculating point defect properties in next generation technologies.

For this reason, calculations using electronic density functional theory
(DFT) have arisen as a reliable route
to explore the dopability of materials at the atomic scale~\cite{freysoldt_rmp, vandewalle_jap, zunger,  peng_gap}.
However, two sources of error in the associated point defect calculations
limit the application of charged defect DFT efforts
in a high-throughput framework.
First, semi-local exchange-correlation approximations
(e.g., generalized gradient approximation (GGA))
can severely
underestimate the band gap so that usage of post DFT methods becomes
pivotal (e.g., GW~\cite{bruneval_GW, northrup_GW, li_GW} and
GGA+U methods~\cite{ldau,TiO2_defects},
and hybrid functionals~\cite{wu2011}).
Second, applying periodic boundary conditions with finite sized defect supercells to model point defects
makes a defect interact with its own images~\cite{freysoldt_rmp, castleton_review},
thus, causing departure from the key assumption made in the
dilute limit formation energy formalism~\cite{freysoldt_rmp, castleton_review}.
In the case of charged point defects, the finite sized supercell assumption also 
introduces the need for correcting the electrostatic 
potential
~\cite{freysoldt_rmp, zunger}.
Typically, the strongest defect-defect
interaction 
is the Coulomb interaction between charged point defects. 
Based on well-known scaling laws,
these interactions were first treated
with computationally costly supercell scaling methods, 
which require multiple calculations for each defect~\cite{castleton_review}.
A faster route to computing defect formation energies 
became available with the development of \textit{a posteriori}
correctional techniques.
While the \textit{a posteriori} corrections allow for fewer calculations to be performed, 
their usage requires experience in 
addressing issues arising from delocalization of the defect wavefunction~\cite{zunger}.
Furthermore,
the calculations are often resource demanding and tedious because of the large number of
pre- and post-processing steps involved.

To address these problems
we have developed the Python Charged Defects Toolkit (PyCDT),
which enables expanded applications
in the context of materials discovery and design.
Our python-based tools automate the setup and analysis of
DFT calculations of isolated intrinsic and extrinsic point defects
(vacancies, antisites, substitutions, and interstitials)
in semiconductors and insulators.
While other efforts have recently been made available with similar objectives~\cite{goyal2016computational, pean2017presentation, yim2017property},
PyCDT is unique in its direct queries to the Materials Project~\cite{jain2013commentary} database (expediting chemical potential and stability analysis
for Perdew--Burke--Ernzerhof (PBE)
GGA calculations)~\cite{pbe_functional}.

A central objective of defects modeling
in non-metallic systems is determining 
the relative stability of different defect charge states.
PyCDT therefore implements the defect formation energy formalism
reviewed in Sections~\ref{introdefectsection},~\ref{thermosection},~\ref{chemicalpotentialsection} and
\ref{supercellsection}. To minimize the errors in defect formation energies arising
from the periodic boundary conditions,
PyCDT supports the commonly used correction scheme due to Freysoldt et
al.~\cite{freysoldt2009} and its extension to anisotropic systems
by Kumagai and Oba~\cite{kumagai2014}
(Section~\ref{supercellsection}). 
Our tools also include charge-state assignment procedures
developed on the basis of extensive literature data
(Section~\ref{chargeassignmentsection})
and an effective interstitial-finding algorithm~\cite{zimmermann2017}
(Section~\ref{interstitialsection}).
Furthermore, PyCDT provides a user-friendly command-line interface
that provides ready access to all tools.
We demonstrate the setup and analysis of the defect calculations from the command line in Section~\ref{sec:execution} 
using gallium arsenide (GaAs) as an example system and by
employing the widely-used \textsc{Vienna Ab initio Simulation Package}
(VASP)~\cite{kresse1993ab,kresse1994ab} as a backend DFT software. 
In Section~\ref{sec::valid}, we validate the finite-size charge correction schemes implemented 
and verify the results obtained for GaAs.
We emphasize 
that our approaches and implementations
are entirely general, thus, seamlessly facilitating extensions
to other DFT packages.

\section{Background and Methods\label{sec:methods}}
\begin{figure}[!ht]
  \begin{center}
    \includegraphics[width=5.5cm]{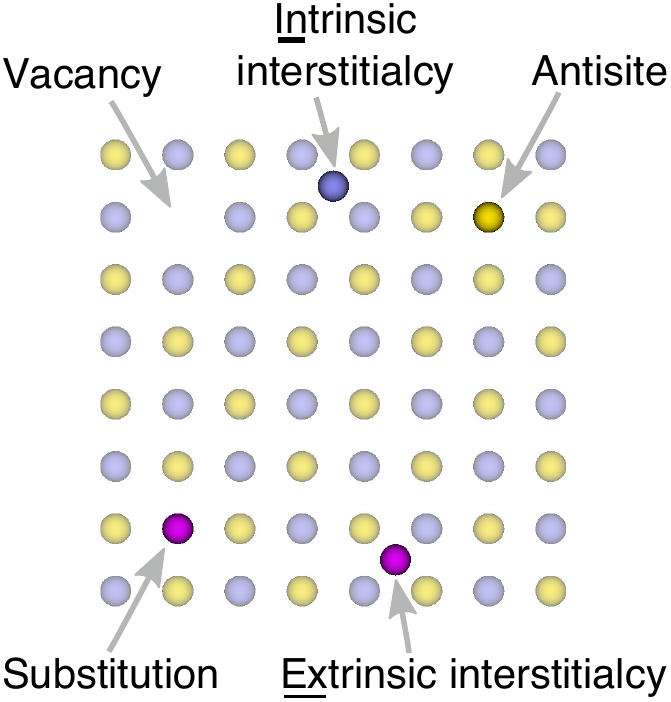}
    \caption{\label{fig:def:types}
      Intrinsic point defects
      (top: vacancy, intrinsic interstitialcy, antisite) and
      extrinsic point defects (bottom: substitution, extrinsic interstitialcy).}
  \end{center}
\end{figure}

In general, point defects can be divided into two categories: intrinsic and extrinsic~\cite{seebauer2006}. 
Intrinsic (or native~\cite{seebauer2006,freysoldt_rmp})
point defects (Figure~\ref{fig:def:types}: top)
involve only chemical species
that are part of the perfect bulk material (e.g., Si
in silicon).
For elemental materials,
there are two basic intrinsic defect types:
vacancies (e.g., vac$_\mathrm{Si}$, denoting a vacancy on Si site)
and self-interstitials (e.g., Si$_{i}$).
For compounds (e.g., GaAs), 
there is an additional defect type: antisites (e.g., As$_\mathrm{Ga}$).
Because intrinsic point defects are equilibrium defects due
to configurational entropy,
they can be well described and their occurrence understood
in the framework of equilibrium thermodynamics (formation energies,
$E^\mathrm{f}$, used to predict equilibrium concentrations, $c$). 

Extrinsic defects (Figure~\ref{fig:def:types}: bottom),
which are also referred to as
impurities~\cite{seebauer2006,freysoldt_rmp},
introduce a foreign chemical species into the perfect bulk
material~\cite{callister2007}.
These include substitutional defects (e.g., Mn$_\mathrm{Ga}$) and
extrinsic interstitials (e.g., Mn$_{i}$).
We distinguish between extrinsic and
intrinsic defects 
because impurities are often inserted on purpose (intentional doping)
under well-defined conditions to achieve desired material properties;
in particular, electrical and optoeletronic properties~\cite{seebauer2006}.
The conditions under which extrinsic defects are inserted
(e.g., via implantation or quenching) 
often differ extremely 
from the thermodynamic equilibrium assumption made in the 
assessment of intrinsic point defects.
Despite the limited conceptual applicability, the thermodynamic framework still represents 
the most commonly pursued route to assessing ``dopability'' of materials ~\cite{freysoldt_rmp,dorenbos2005,chaudhry2014,zunger_TCO, scanlon_TCO, varley_TCO,zakutayev_solar, walsh_solar,zhu_te, pomrehn_te}. 

In contrast to metals,
point defects in semiconductors and insulators
can carry a charge~\cite{seebauer2006} 
localized around the defect
site. 
Physically, these 
charged point defects
introduce 
states within the band gap which can trap 
charge carriers (electrons and holes).
Defect states that are close to
the band edges are able to
ionize to create free carriers, while states that are deep in the gap lead to strong carrier trapping.
This may be wanted (e.g., in photovoltaics~\cite{walsh_solar})
or not (e.g., solid-state electrolyte batteries~\cite{ss_electrolytes}).
Because many technological applications
use intentional doping to improve performance,  
knowledge of the capacity to dope a material (``dopability'') is desirable
and motivates the exploration of defect properties with theoretical methods.

\subsection{Formalism for equilibrium point defects}\label{introdefectsection}
The thermodynamics of point defects has been the subject of many
excellent reviews (see, for example,
refs.~\cite{freysoldt_rmp, vandewalle_jap, zunger}\ 
and references therein),
which have presented and discussed the underlying physics
and properties in great detail.
In the following sections, we focus on describing the procedures implemented in PyCDT
for computing quantities of interest for point defects (i.e.,
formation energies and transition levels) with DFT calculations.
The applications of the defect formalism to be described are limited by the intrinsic
shortcomings of DFT
(e.g., the well-known underestimation of the band gap~\cite{zunger}).

 There is a hierarchy of DFT-based methods that can be used within the defect formalism
implemented by PyCDT.
The simplest approximation in DFT is the use of a semilocal functional
(i.e., local-density approximation (LDA), GGA),
which is computationally most efficient,
but has well-known limitations due to band gap inaccuracies.
Higher levels of theory include hybrid-functionals and meta-GGA,
both of which can be used to achieve higher accuracy,
however, at an increased computational cost~\cite{metaGGASCAN, hybridreview}.
Despite the limitations of semilocal DFT, defect calculations 
have proven useful for revealing the dominating defects
under different growth conditions encountered in many experiments, such as the III-V semiconductors~\cite{tahini}.
While recent developments in hybrid functionals and meta-GGA have shown 
promise in addressing the inherent limitations in accuracy associated with 
semi-local functionals~\cite{metaGGASCAN, hybridreview}, 
recent evidence shows that new improvements to the approximations for exchange and correlation in one system do not always yield universal improvements for other systems with similar chemistries~\cite{medvedevscience2017}. With this fact considered, semi-local approximations at least have the benefit of having predictable errors which can be corrected with appropriate techniques~\cite{zunger}.

While PyCDT's unique interface with the Materials Project (MP) database, which is composed of GGA and GGA+U level data, suggests a restriction to semi-local approaches, PyCDT has many functionalities which help 
place defect formation energetics 
closer to 
those obtained from higher levels of theory.
One feature that is particularly useful is the ability for PyCDT to help expedite the setup and parsing stages of defect calculations performed on higher levels of DFT theory (\emph{e.g.} for improved chemical potentials, setting up a user's personal phase diagram calculation based on a composition of interest). These features are described further in the following sections.

\subsection{Defect Formation Energies}\label{thermosection}

\begin{figure}[!ht]
  \begin{center}
    \includegraphics[width=15cm]{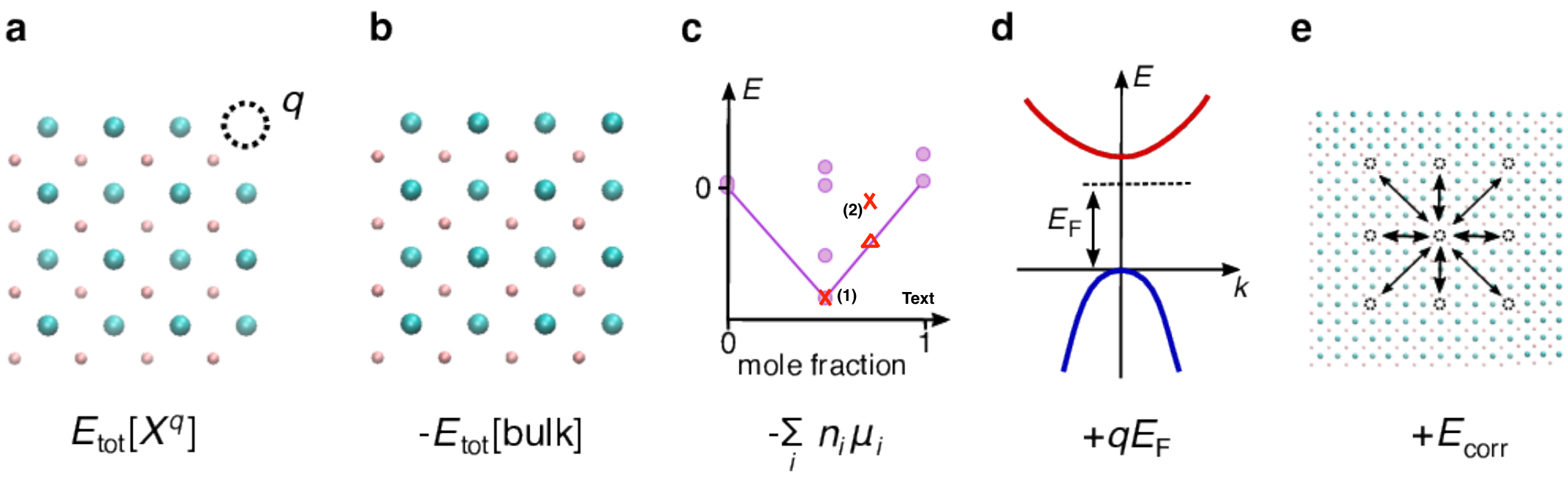}
    \caption{\label{fig:Eformation}
      Different contributions to the formation energy:
      (a) energy of defective supercell in charge state $q$,
      (b) energy of pristine bulk supercell,
      (c) atomic chemical potential computed from the ground state hull,
      (d) electron/hole chemical potential generated from electron reservoir, and
      (e) correction terms to account for defect-defect interactions
          arising from periodic boundary conditions
          as well as for the homogeneous background charge
          which requires potential re-alignment. The different labeled X's given in \ref{fig:Eformation}c show stable and metastable compounds. The latter case yields a complication which will be discussed in Section \ref{chemicalpotentialsection}}.
  \end{center}
\end{figure}

The primary quantity of interest is the formation energy, $E^\mathrm{f} [X]$, which is the
energy cost to form or create an isolated defect, $X$, in a bulk or host material.
The formation energy of an isolated defect (i.e., in the dilute limit)
depends on the defect charge state, $q$: $E^\mathrm{f} [X^q]$. It can be calculated from DFT supercells using:

\begin{equation}
E^\mathrm{f} [X^q] = E_\mathrm{tot}[X^q] - E_\mathrm{tot}[\mathrm{bulk}] - \sum\limits_{i}n_i\mu_i + qE_\mathrm{F} + E_\mathrm{corr}
\label{eqn:formen}
\end{equation}
We illustrate this equation graphically in Figure \ref{fig:Eformation}
and note that each term will be described in detail in subsequent sub-sections.
$E_\mathrm{tot}[X^q]$ and $E_\mathrm{tot}[\mathrm{bulk}]$ are the total DFT-derived energies
of the defective and pristine bulk supercells, respectively.
The third term, $-\sum_{i}n_i\mu_i$, is a summation over the atomic chemical potentials, or the energy cost of an atom, $\mu_i$, being added ($n_i=+1$) or removed ($n_i=-1$) from the bulk undefective supercell.
The atomic chemical potential can reflect the growth conditions of the material,
allowing this formalism to be used to guide defect engineering approaches 
(cf., Section \ref{chemicalpotentialsection}).
The fourth term, $qE_\mathrm{F}$, represents the 
energetic cost of adding or removing electrons,
where $E_\mathrm{F}$ is the Fermi energy, which serves as the chemical potential of the electron reservoir. 
The Fermi energy is usually referenced to the valence band maximum
from a band structure calculation,
such that the formation energy can be plotted
as a function of the Fermi energy across the band gap.  
Finally, $E_\mathrm{corr}$ is a correction term due to the presence
of periodic images
that becomes necessary for charged defects in DFT supercell calculations.
This correction has drawn significant attention from the defects modeling  community,
resulting in a number of alternative computational approaches that are
discussed in more detail in Section \ref{supercellsection}.

\subsection{Chemical Potentials} \label{chemicalpotentialsection}
The atomic chemical potential is associated with the thermodynamic energy cost 
for exchanging atoms between the defect and a thermodynamic reservoir. 
The individual chemical potentials are set by the composition of the 
material (e.g., the mole fraction of As in GaAs) which itself is determined by the defect
formation energies. Hence, two approaches can be used to derive the 
individual chemical potentials. One involves the use of a statistical 
thermodynamic formalism (\emph{canonical} ensemble approach), 
to compute the concentration dependent free energy 
of the compound and to derive the relationship between the chemical potentials and 
composition~\cite{wolvertonNainPbTe}. More frequently, bounds on the chemical potentials 
are set (\emph{grand-canonical} ensemble approach), in a manner first defined by Zhang and 
Northrup~\cite{zhang}, from zero-temperature energies alone.
Only the grand-canonical approach is currently included in PyCDT,
whereas future versions will also support canonical approaches 
to chemical potential calculations.
Below we demonstrate the grand-canonical
approach 
for the simple example of GaAs. 

The bulk energy (or free energy at finite temperature)
per formula unit, $\mu_{GaAs}^0$, fixes a relation for
the chemical potentials of gallium, $\mu_{Ga}$, and 
arsenic, 
$\mu_{As}$, respectively:  
$\mu^0_{GaAs} = \mu_{Ga} + \mu_{As}$.
For Ga-rich compositions, the values of $\mu_{Ga}$
are constrained by stability of the compound relative to the precipitation 
of excess Ga to form a bulk Ga phase. At zero temperature, this constraint can be expressed as 
$\mu_{Ga} < \mu_{Ga}^0$, where $\mu_{Ga}^0$ is the energy per atom of bulk Ga. Thus, one 
extremum
can be selected as the ``Ga-rich'' limit, where 
$\mu_{Ga}=\mu^0_{Ga}$. The atomic 
chemical potential of As is then fixed: $\mu_{As}=\mu^0_{GaAs}-\mu^0_{Ga}$.
The same 
approach holds for As-rich compositions
(above:
interchange Ga and As labels with each other). 
While the Ga-As system has a phase diagram with just one unique compound,
the more general case has multiple stable compounds, which requires the limits of stability to be expressed in terms of the formation of compounds with neighboring compositions within the phase diagram.
In general, the chemical potentials in an n-component system will be defined in PyCDT by defining the limits of stability for the different possible n-phase states of equilibria.

As an example, consider the Sn-Se system which has a 0~K\ ground-state hull that
 contains the phases Sn, SnSe, SnSe$_2$ and Se. When calculating defects in the SnSe phase,
  the ``Se-rich'' limit would instead be defined by equilibrium with the SnSe$_2$ phase (Equation \ref{snse2eqn}), 
combined with the 
stability condition for bulk SnSe (Equation \ref{snseeqn}). 
This forms a
system of equations for the chemical potentials of $\mu_{Sn}$ and $\mu_{Se}$. 
\begin{align}
\mu^0_{SnSe_2} & >\mu_{Sn} + 2\mu_{Se}\label{snse2eqn}\\
\mu^0_{SnSe}&=\mu_{Sn} + \mu_{Se}\label{snseeqn}
\end{align}

This formalism 
for calculating equilibrium bounds on the chemical potentials
requires knowledge of the ground-state hull,
which governs the zero-temperature limit of the phase diagram of the system.
The advantage of this formalism is that defect formation 
energies can be obtained entirely from first principles calculations,
rendering experimental input unnecessary.
The drawback is that one must compute the full phase 
diagram of the system using the same functional choice used for the defect 
calculations. In applications of the PyCDT code based on the
PBE-GGA exchange-correlation potential, 
PyCDT uses the ground-state hulls that are made available
through the 
MP database~\cite{jain2013commentary}.
PyCDT has integrated functionality that 
queries the MP database for every computed DFT entry in the phase   
diagram so that no new calculations are required to compute the bounds on the 
chemical potentials.
Note that correct usage of the MP data for defect calculations requires
consistency between personal defect calculations and the calculations from the MP. 
This is easily checked through the compatibility tools available in pymatgen~\cite{ong2013python}.
If the bulk phase is thermodynamically 
stable and is not already computed in the MP database, PyCDT manually 
inserts the computed phase into the phase diagram and then provides all of the associated
bounds on the chemical potentials. 
If a user prefers to compute chemical potentials on a different level of theory than is provided by the MP, 
the core code of PyCDT can be used to setup and calculate atomic chemical 
potentials 
through first pulling the composition's phase diagram, and using the structural information to setup a personal phase diagram calculation for the user. In a similar manner, this information can be setup 
for any DFT code desired by the user through the use of Pymatgen's code agnostic classes.

For highly correlated systems such as transition metal oxides, MP settings default to GGA+U. 
When computing the phase diagram that contains a 
mixture of GGA and GGA+U computed phases, MP employs the mixing scheme of 
Jain et al.~\cite{jain2011b}, which adds an empirical correction to the energies of GGA+U compounds. The mixing 
scheme was shown to give formation energies that are consistent with 
experimental data with a mean absolute relative error of under 2\%. 
The resulting correction term to the chemical potentials was found to be important in several defect studies~\cite{medasani2017,tahini2016}.

One complication that should be mentioned 
is the case where the compound under consideration
does not reside on the convex hull.
That is, the compound is higher in energy than another compound with the
same composition or with respect to phase separation to compounds with other compositions.
In this instance, the calculation is predicting 
the compound to not be present in the equilibrium phase diagram in the limit of zero temperature.
For small energy-above-hull values, this situation could indicate that the compound is stabilized 
by entropic contributions at finite tempearture~\cite{agoston_vibentropy}, or could be an artifact of the previously mentioned inaccuracies of DFT~\cite{hautier_dftenergies}, or the 
experimentally synthesized phase exists in a state of metastable equilibrium~\cite{metastable_gerd}.
Regardless of the reason, a positive energy-above-hull value presents a practical problem for defining the chemical potential.

In such cases,
PyCDT issues a warning and
the chemical potentials used are
with respect to the phases in equilibrium at the given composition in the phase diagram.
In figure \ref{fig:Eformation}c, this situation is graphically represented by the data points labeled ``(2)", where the red X above the hull is the compound of interest and PyCDT uses the red triangle to define a set of compounds in equilibrium for defining the atomic chemical potentials.
In these instances, the computed defect physics should 
be interpreted with caution.
The user may wish to define the chemical-potential limits based on more detailed 
knowledge of the growth conditions or
compute the chemical potentials from a full finite-temperature 
free energy model of the compound of interest.

\subsection{Periodic Supercell Corrections}\label{supercellsection}

Periodic boundary conditions (PBCs) are the standard way
to deal with the regular arrangement of crystalline solids in DFT calculations.
Once a defect is introduced, PBCs can give rise to sizable 
interactions
of the defect with its periodic images,
contrasting the 
assumption made above for the dilute-defect limit. 
Because this limit is consistent
with the thermodynamic formalism outlined in Sec.~\ref{thermosection},
the interactions between neighboring 
defect images should be minimized to yield accurate formation energies.
For charged defects in semiconductors and insulators,
Coulombic interaction with neighboring images exists, which decays as 1/L, where L is the supercell periodic length.
The charge interactions are the 
dominant effect that need to be taken into account when correcting the 
formation energy of defects in non-metals.
Elastically-mediated interactions
which are due to the strain fields induced
when the positions of atoms near the defects 
also exist, but decay more rapidly in real space, and are often minimal~\cite{freysoldt_rmp}. In cases where these interactions are important, methods have been developed to account for them (see ~\cite{freysoldt_rmp} and references therein), which will not be addressed in the following discussion.

To account for the charge correction 
one approach has been
to create successively larger defect supercells.
Scaling laws for the 
electrostatic interactions with respect to system size
are then used
in order to extrapolate $E^\mathrm{f}$ to the dilute limit \cite{lany_fs, taylor_fs}. 
An alternative approach is based on an \emph{a posteriori} analysis of the electrostatic potential
for a single supercell calculation~\cite{freysoldt2009, freysoldt2011}.
An important requirement for the alternative approach
is that the charge be sufficiently localized within the vicinity of the defect.
If so, a moderately sized defect supercell typically suffices,
hence, offering a computationally more efficient route to
calculating reliable defect formation energies.
The latter methodology, referred to as ``correction methods'' in the following,
is employed in PyCDT.

Correction methods address two issues:
\begin{enumerate}
\item the electrostatic energy
from the interaction between the charged defect and its images, and 
\item a potential alignment term that corrects for a 
fictitious jellium background required to maintain overall charge neutrality in the system.
\end{enumerate}
Many different methods have been proposed to correct for these two terms, 
as summarized in several comprehensive reviews (see, e.g., ~\cite{freysoldt_rmp, vandewalle_jap, zunger, castleton_review} and references therein). 

The theoretical starting point for the correction methods 
considers a periodic array of point charges (cf., Figure~\ref{fig:Eformation}e)
with an associated Madelung energy, $E_\mathrm{M}$:
\begin{equation}
E_\mathrm{M} = \frac{q V_\mathrm{M}}{2} = \frac{q^2 \alpha}{2 \epsilon L}
\end{equation}
where $V_\mathrm{M}$ is the Madelung potential,
$\epsilon$ is the dielectric constant, and $\alpha$ is the Madelung constant
which solely depends on the geometry of the periodic array. 
Makov and Payne~\cite{MakovPayne_correction}
introduced one of the earliest charge correction methodologies
by deriving the next leading order term to the interaction potential.
This results in a term that scales as $L^{-3}$,
and, therefore, most supercell scaling approaches
fit uncorrected formation energies to the form of $a L^{-1}+ b L^{-3}$. 
Komsa \emph{et al.}~\cite{komsa2012} used this supercell scaling method 
for evaluating the performance of different correction methods
that are based on single supercell calculations. 
They concluded
that the correction by Freysoldt~\emph{et al.}~\cite{freysoldt2009}
produces the most reliable charge corrections
for defects with charges that are well localized within the supercell.
From all considered defects,
the authors calculated a mean absolute error of
0.09~eV\ in the formation energy
between the estimate from the 64-atom supercell with charge corrections
and the estimate from the supercell-scaling method (i.e., using
extrapolation toward the dilute-defect limit, but without
applying any charge correction).

PyCDT includes a Python implementation of the 
correction scheme derived by Freysoldt \emph{et al.}~\cite{freysoldt2009} and implemented  
in the open-source DFT software S/PHI/nX~\cite{boeck2011}.
The approach is based on a separation of the long-range and 
short-range interactions between charged defects, using information 
directly outputted from a DFT calculation. 
Originally, 
an isotropic dielectric constant was assumed.
Recently, Kumagai and Oba~\cite{kumagai2014}
extended the approach to anisotropic systems,
by considering the full dielectric tensor.
The analytic expression of the Madelung potential under isotropic 
conditions facilitates the use of a Gaussian
distribution for the defect charge,
whereas the analytic expression of Madelung potential for 
anisotropic systems is limited to point charges.

The authors of the two correction methods suggest different approaches to 
calculating the potential alignment correction. The isotropic correction by Freysoldt \emph{et al.}
uses a planar average of the electrostatic short range potential while the 
anisotropic correction by Kumagai and Oba takes averages of this same 
potential at each atomic site outside a given radius from the defect. 
Both approaches are available in the PyCDT code, with the 
isotropic correction by Freysoldt \emph{et al.} being the default.
The planar averaging method can become problematic when large relaxation occurs,
as the atomic sites 
contribute heavily to the change in electrostatic potential. The atomic site averaging method can 
become problematic if a small cell size results in a small number of atoms being 
sampled, causing statistical sampling errors. 
While, in principle, these alignment corrections should be equivalent,
tests that we conducted revealed non-negligible discrepancies.
However, the potential-alignment term often tends to be small
($\sim$0.1 eV), and, therefore, 
to not change overall trends in defect formation energies.

The correction methods used for point defects in semiconductors and insulators
have been an intensely debated topic in the past decade~\cite{freysoldt_rmp}.
One issue upon which there is common agreement is
that large defect-defect interactions change the energetics of the system
so that the computed defect formation energies
are no longer relevant for physical quantities like defect concentrations or 
thermodynamic transition levels. 
These unwanted defect-defect interactions frequently lead
to delocalization of the defect charges,
the instance of which has to be ascertained manually.
Several methods that address delocalization can be found
in the literature~\cite{freysoldt_rmp,zunger, lany_fs}.

If we assume that the defect charge can indeed be localized
within the level of DFT used, then best practice demands
to balance computational expediency (supercell size)
with sufficient localization of the charge around a defect
as indicated by the outputs of the charge correction method chosen.
In the original derivation of the isotropic correction by Freysoldt \emph{et al.},
the middle ``plateau" region of the electrostatic potential
yields information about the
separation of long range and short range effects.
A flat plateau indicates
that the Coulomb potential has been removed from the total potential generated by DFT, 
and short range effects have not delocalized throughout the entire supercell.
As a result, Freysoldt \emph{et al.}~\cite{freysoldt2009} suggested
that the flatness of the resultant ``plateau'' yields a qualitative metric
for the success of the calculation.
When running the isotropic correction by Freysoldt \emph{et al.} in PyCDT,
the planar averaged electrostatic potential is analyzed
for variations larger than 0.2~eV---a number that stems from experience,
and can be altered in the code if the user prefers to. 
If this criterion is not met, the code raises a warning.
In such a case, the user should consider
the possibility of delocalization.

For users who desire additional corrections related to improving corrections with semilocal functional approaches, the development branch of PyCDT also includes the ability to include band edge level alignment with respect to the average electrostatic potential, shallow level corrections based on the values for band edge alignment, and Moss-Burnstein band filling corrections~\cite{zunger}. Parts of these corrections require some subjective judgment calls to be made by the user, so they are not included in the automation procedure by default.

\subsection{Charge Ranges}\label{chargeassignmentsection}

Charge ranges, $[q_\mathrm{min}, q_\mathrm{max}]$,
have to be estimated beforehand for a given defect $X$,
and for this purpose ionic models typically form the basis of such
predictions~\cite{petretto2015}.
Known oxidation states of the element(s)
involved in $X$ can then be used to 
define the charge states to be considered.
However, common ionic models do not always predict the 
most stable defect.  Tahini \emph{et al.}\ have, in this context, shown that
combining gallium or aluminum with group-V elements can
yield negatively charged anion vacancies,
whereas an ionic model predicts a $+3$ charge state \cite{tahini}.
In PyCDT, we implemented different procedures to determine the 
range of defect charges for semiconductors and insulators.
Users can choose between either of these two options and
a custom range of defect charges for each defect,
as described in Section~\ref{setupcalcs}.

To address the issue of uncommon charge states found in semiconductors, we developed
a data-driven approach that combines elemental oxidation states
with results from literature
for determining the optimal charge assignment process.
We compiled a list of stable charge 
states (Table~\ref{tab:charge:states}) 
from previous studies 
for various defects in
zinc blende and diamond-like semiconductor structures~\cite{wu2011, tahini, petretto2015, deak,
chroneos,corsetti,li2012,dossantos2011,neugebauer1994}.
Procedures adhering more or less
strictly to ionic models resulted in too few
charge states when compared with the literature.
The most effective approach
that we found
employs a bond-valence estimation scheme
\cite{ong2013python, okeeffe1991} to obtain formal charges 
of elements in the bulk structure,
as well as minima and maxima of common oxidation states of bulk and defect elements. 
The formal charges and common oxidation-state ranges are subsequently
used in a defect type-dependent assignment procedure:
\begin{enumerate}
\item \emph{Vacancies}: Use the formal charge of the species originally located on the vacant site,
                        $oxi$, to define the charge range: $[-oxi, +oxi]$. For \ce{GaAs}, this procedure 
results in defect charges ranging from -3 to 3 for both $\mathrm{V}_\mathrm{Ga}$ and $\mathrm{V}_\mathrm{As}$.
\item \emph{Anti-sites}: Use the minimum and the maximum from combining all oxidation states
                         of \emph{all} elements in the bulk structure, $oxis_\mathrm{bulk}$,
                         to define the relevant charge range:
                         $[\mathrm{min}(oxis_\mathrm{bulk}),\mathrm{max}(oxis_\mathrm{bulk})]$.
                         Data mining determined that the upper range limit can, in fact, be decreased by
                         2: $[\mathrm{min}(oxis_\mathrm{bulk}),\mathrm{max}(oxis_\mathrm{bulk})-2]$.
With this procedure, the antisites in \ce{GaAs} are assigned charge values from -3 to +3. 

\item \emph{Substitutions}: Determine the oxidation states of the foreign (or, extrinsic) species, 
                              $oxis_\mathrm{ex}$. Then subtract the formal charge of the site species
                              to be replaced from this list. Use the minimum and maximum of this set to produce
                              $[\mathrm{min}(oxis_\mathrm{sub}),\mathrm{max}(oxis_\mathrm{sub})]$.
                              Data mining determined that, 
                              if the new range has more than 3 charge states and has an upper bound larger than 2,
                              one can cap the range by 3 to prevent excessively high charge states. For example, 
when GaAs is doped with Si, $\mathrm{Si}_\mathrm{Ga}$ generates charges in the range of [-7,1], and  
$\mathrm{Si}_\mathrm{As}$ generates charges in the range of [-1,4].
\item \emph{Interstitials}: Use the minimum and maximum of all oxidation states of the interstitial species: \\
                            $[\mathrm{min}(oxis_\mathrm{int}),\mathrm{max}(oxis_\mathrm{int})]$.
                            If 0 is not included, data mining suggests that we extend the range to 0 accordingly.
For As interstitials in GaAs, the resulting defect charges are in the range [-3, 5].
\end{enumerate}

The algorithm successfully includes all charge states from our benchmark
list in Table~\ref{tab:charge:states} by yielding, on an average, 6.4
~states\ per defect.
The average number of states produced too much per defect
in comparison to literature
(excess charge states) are 1.1  
is 1.9 at the lower (more negative) and
at the upper (more positive) charge bound, respectively.
This is desirable because 
including more charge states on either side 
ensures no extra states become 
stabilized when varying the Fermi level within the band gap.
Hence, the effective relative excess in charge states is
$20\%$ 
and, thus, acceptable.

For insulators, the number of defect charge states
is typically less than for the above discussed semiconductors.
For example, the charge states in \ce{MgO} range from
$-2$ to $0$ and $0$ to $+2$ for cation and anion vacancies, 
respectively~\cite{gibbson1994}. 
Any other charge state is not considered because of the high 
ionization energy required to form \ce{Mg^{3+}} and
the high electron affinity of \ce{O^{2-}} to form \ce{O^{3-}}.
Hence, the oxidation states of cations and anions
are limited to $[0,y]$ and $[-x,0]$
for a binary $\mathrm{A}_x \mathrm{B}_y$ insulator,
where A is a cation and B is an anion.

\subsection{Interstitials}\label{interstitialsection}

PyCDT uses an effective and easily extendable
approach for interstitial site finding
(Interstitial Finding Tool: InFiT)
that has been recently introduced by
Zimmermann \textit{et al.}~\cite{zimmermann2017}.\ 
The procedure systematically searches for tentative interstitial sites
by employing coordination pattern-recognition
capabilities~\cite{peters2009,zimmermann2015}
implemented in pymatgen~\cite{ong2013python}.
In Algorithm~\ref{algo:inter},
we provide a simplified pseudo-code representation
of the approach.
The detected interstitial sites exhibit coordination patterns
that resemble basic structural motifs (e.g., tetrahedral
and octahedral environments).
Such interstitial sites
are particularly important
because several~\cite{Decoster:ApplPhysLett:2008,
Decoster:PhysRevLett:2009, Decoster:JApplPhys:2009,
Decoster:PhysRevB:2010, Decoster:ApplPhysLett:2010,
Pereira:ApplPhysLett:2011, Pereira:PhysRevB:2012,
Decoster:JApplPhys:2012, Amorim:ApplPhysLett:2013,
Silva:JApplPhys:2014}
$\beta ^-$ emission channeling
measurements~\cite{Hofsass:PhysRepRevSecPhysLett:1991,
Silva:RevSciInstrum:2013}
have identified them
as the most prevalent types of isolated
defects after substitutions
for impurities implanted
into similar materials as we consider here
(zinc blende/wurtzite-like and diamond-like
structures).
Bond-center interstitials
in a so-called split-vacancy configuration
are also observed frequently.
However, these are defect complexes---not isolated
defects---and, thus, beyond the scope of the present
PyCDT implementation.
The interstitial search approach
should also be suitable for intercalation and ion
diffusion applications because related design
rules typically rely on detection of tetrahedral,
octahedral, bcc-, and fcc-like
environments~\cite{Rong:ChemMater:2015,
Wang:NatMater:2015}.

\begin{algorithm}
\caption{Interstitial Site Searching}
\label{algo:inter}
\begin{algorithmic}[1]
    \State ${GetNeighbors({struct},\ {inter\_trial\_site},\ {dmin},\ {delta})}$ \Comment Get a list of neighbors of site ${inter\_trial\_site}$ in ${struct}$ that are within a sphere of $(1+{delta})\;dmin$.
    \State ${GetCoordinationDescriptor(point,\ neighs,\ coord\_type)}$ \Comment Calculate the value of the descriptor for the target coordination environment ${coord\_type}$ of an atom at ${point}$ with neighbors ${neighs}$.  
    \State

    \Procedure{GetCoordinationPatternInterstitials}{${struct}$, ${inter\_elem}$, ${all\_coord\_types}{=[\mathrm{tet}, \mathrm{oct}]}$, ${thresh}=[0.3, 0.5]$, ${dl}=0.2$, $dstart=0.1$, $ddelta=0.1$, $dend=0.7$}
    \State Let ${inter\_sites}$ be a new empty list
    \State Let ${coord\_descr}$ be a new empty list
    \For{${point}$ on a regularly mashed grid in ${struct}$ with resolution ${dl}$}
      \State Let ${dmin}$ be the distance from ${point}$ to the closest crystal atom
      \If{${dmin} > 1$~\AA}
        \State Let ${inter\_trial\_site}$ be a new ${Site}$ object of type ${inter\_elem}$ located at ${point}$
        \State Let ${struct\_plus\_inter}$ be a new ${Structure}$ object $\gets {struct}$ appended by ${inter\_trial\_site}$
        \For{${delta}$ starting from $dstart$ in $ddelta$ steps to $dend$}
          \State Let ${neighs}$ be a new list $\gets {GetNeighbors({struct\_plus\_inter}, {inter\_trial\_site}, {dmin}, {delta})}$
          \For{${coord\_type}$ having index ${icoord}$ in ${all\_coord\_types}$}
            \State Let ${this\_coord\_descr}$ be a float number $\gets {GetCoordinationDescriptor(point, neighs, coord\_type)}$
            \If{${this\_coord\_descr} > {thresh[icoord]}$}
              \State Add entry to ${inter\_sites}$ list $\gets {inter\_trial\_site}$
              \State Add entry to ${coord\_descr}$ list $\gets {this\_coord\_descr}$
              \State \textbf{break}
            \EndIf
          \EndFor
        \EndFor
      \EndIf
    \EndFor
    \State Let ${labels}$ be a list of ${coord\_type}$-specific cluster labels for each entry in ${inter\_sites}$ which are found with a distance threshold of $1.01\ {dl}$.
    \State Let ${include}$ be a new empty list.
    \For{${unique\_label}$ in ${labels}$}
      \State Find site with index ${imax}$ in ${inter\_sites}$ that has highgest ${coord\_descr}$ among site with this ${unique\_label}$.
      \State Add entry to ${include}$ $\gets {imax}$
    \EndFor
    \State ${final\_inter\_sites}$ $\gets $ Prune the ${inter\_sites}$ in ${include}$ further to include symmetrically distinct sites only
    \State \Return ${final\_inter\_sites}$
    \EndProcedure
\end{algorithmic}
\end{algorithm}

\subsection{Default DFT Calculation Details}
PyCDT includes mechanisms to input user-defined settings for all DFT calculations. 
If no user settings are specified, then the following initial settings are specified for the calculation.
The ions in the defective supercells are geometrically relaxed
at constant volume until the force on each ion is less than 0.01~eV/\AA.
At each geometric step, the energy of the supercell is converged
to 10$^{-6}$~eV.
By default, spin polarization is turned on, and 
crystal symmetry is ignored to account for any symmetry breaking
relaxations such as Jahn-Teller distortions.
Electronic states are populated using a Gaussian smearing 
method~\cite{devita1992} with a width of 0.05 eV.
While no geometric relaxation is performed for the non-defective 
bulk supercell, a calculation that is necessary for the charge correction, 
the electronic degrees of freedom are optimized
with the same settings applied to the defect-supercell calculations.
A $2\times2\times2$ Monkhorst-Pack $k$-point mesh is
used for the defect calculations.
For any other parameters, we adopt the standard MP settings~\cite{ong2013python}.
Finally, we emphasize
that PyCDT also includes mechanisms 
to input user-defined settings for DFT calculations which can be used for extending calculations beyond the exchange
and correlation approximation of GGA.

\section{PyCDT Usage and Examples} 
\label{sec:execution}

\begin{figure}[!ht]
  \begin{center}
    \includegraphics[width=13cm]{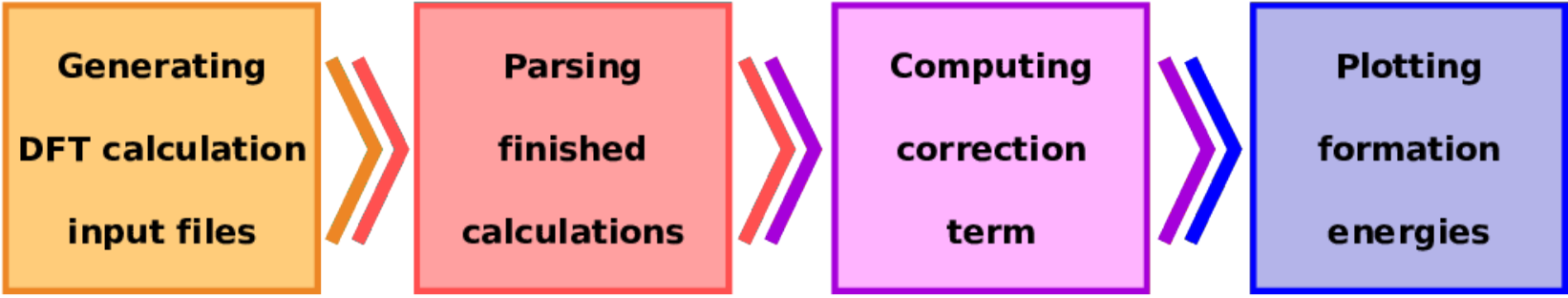}
    \caption{\label{fig:anal:steps}
      Steps in the computation of charged-defect formation energies
      with PyCDT.}
  \end{center}
\end{figure}

PyCDT was developed in a way
that reflects different analysis stages
(Figure~\ref{fig:anal:steps}):
setup of DFT calculations,
parsing of finished jobs,
computation of a correction term,
and plotting of formation energies.
This allows reuse and integration of parts of the code
in other packages.
Our package has a dependency on pymatgen~\cite{ong2013python}, 
matplotlib~\cite{matplotlib} and numpy~\cite{numpy},
and it was developed and tested for Linux and Mac OS X.
However, we also expect it to work on Windows (with cygwin).

Below, we describe in detail
how to proceed through the general charged-defect calculation workflow
outlined in Figure~\ref{fig:anal:steps}
using our PyCDT command-line feature.
We demonstrate its usage on the basis of GaAs
and with VASP~\cite{kresse1993ab,kresse1994ab} as the backend
DFT code.
While the command-line interface is a quick and easy route to use PyCDT, 
we point out that the core classes within PyCDT are entirely general and 
can be used for any desired applications that involves setting up, computing charge corrections,
and/or parsing defect calculations.

\subsection{Setup of DFT Defect Calculations}\label{setupcalcs}
The starting point for setting up charged defect calculations
is the crystal structure. 
The user can provide the bulk structure in one of two ways:
\begin{enumerate}
\item by the name of a structure file of conventional format
    (e.g., cif, cssr), or code specific formats such as POSCAR that are recognized by pymatgen, or
\item via a Materials Project identifier (MPID).
\end{enumerate}
Crystal structures from MP~\cite{jain2013commentary}
are obtained through the 
Materials Application Programming Interface (MAPI)~\cite{ong2013python}.
In the MP database, each structure is assigned a unique
identifier.
These MPIDs have the format \textsl{mp-XXX}
in which \textsl{mp-} is prefixed
to a positive integer \textsl{XXX}.
In the following, we use GaAs (\textsl{mp-2534}),
which has the zinc-blende structure,
as an example for performing all different stages of
charged defect-property calculations with PyCDT.

We first generate the defect supercells and the bulk supercell
using pymatgen's defect structure generator and the defect structure classes 
in the core of PyCDT. 
The two steps for generating the input files are combined into a single command:
\begin{framed}
\begin{table}[H]
\begin{tabular}{lllll}
$\triangleright$ pycdt generate\ul{\ \ }input & ( &-{}-structure\ul{\ \ }file & $\langle$structure file$\rangle$ \;\;\;$|$  \;\;\; -{}-mpid $\langle$mpid$\rangle$&)\\
							       & [ &-{}-mapi\ul{\ \ }key &$\langle$mapi\ul{\ \ }key$\rangle$&] \\
							       & [& -{}-nmax &$\langle$max\ul{\ \ }no\ul{\ \ }atoms\ul{\ \ }in\ul{\ \ }supercell$\rangle$&]\\

\end{tabular}
\end{table}
\end{framed}
With \textsl{{-}{-}nmax},
the user defines the maximal number of atoms in the defect supercell.
If the parameter is not given, a default value of 128 is used, which was shown to result in  well 
converged defect formation energies after finite size corrections 
were included in systems with dielectric constants greater than 5.0~\cite{wu2011,komsa2012}. 
The mapi$\_$key 
is required if querying the MP database, and is found on the 
Dashboard after logging into the MP website.

The input file-generation command creates a folder,
representing the reduced chemical formula
of the crystal structure (e.g., $GaAs$).
It contains several subfolders,
whose names are indicative of the calculations to be performed:
\begin{itemize}
\item \textsl{bulk}: calculation of pristine crystal structure,
\item \textsl{dielectric}: calculation of macroscopic static dielectric tensor 
    (ion clamped high frequency, $\epsilon_\infty$, plus the ionic contribution, 
$\epsilon_{ion}$) from DFT perturbation theory, 
(used by the charged defect correction)
\item \textsl{deftype\ul{\ \ }n\ul{\ \ }info}: calculation
      of the $n$-th symmetrically distinct
      defect of type \textsl{deftype}
      (vacancies: \textsl{vac}; antisites: \textsl{as};
      substitutions: \textsl{sub}; interstitials: \textsl{inter})
      with properties \textsl{info}.
\end{itemize}

By default, the above command generates vacancy and antisite defects only.
Hence, there are four defect folders for GaAs,
which has two sublattices,
corresponding to one antisite and one vacancy defect on each sublattice.
Table~\ref{tab:defecttype:foldername}
summarizes the default defect types and
resulting folder names for GaAs.

\begin{table}[H]
    \centering
\caption{\label{tab:defecttype:foldername} Default defects set up for GaAs with PyCDT}
\begin{tabular}{ll}
   \hline
   Defect Type  & Folder Name \\
   \hline
$Vac_{Ga}$  &  GaAs/vac\ul{\ \ }1\ul{\ \ }Ga \\
$Vac_{As}$  &  GaAs/vac\ul{\ \ }2\ul{\ \ }As \\
$Ga_{As}$   &  GaAs/as\ul{\ \ }1\ul{\ \ }Ga\ul{\ \ }on\ul{\ \ }As \\
$As_{Ga}$   &  GaAs/as\ul{\ \ }2\ul{\ \ }As\ul{\ \ }on\ul{\ \ }Ga\\ 
\hline
\end{tabular}\\
\end{table}

Substitutional and interstitial defects have to be
invoked explicitly with the keyword
\textsl{{-}{-}sub host\ul{\ \ }species 
substitution\ul{\ \ }species}.
Multiple  substitutional defects can be generated by repeating the 
\textsl{{-}{-}sub
} keyword with the desired host species and the 
corresponding substituting species.
The substitution folders are labeled in the same manner as the
antisites, only changing \textsl{as} to \textsl{sub}.

The setup of coordination-pattern resembling interstitials
is invoked by the \textsl{{-}{-}include\_interstitials}
command-line option.
PyCDT produces intrinsic interstitials as per default only.
Extrinsic interstitials can be achieved by providing a
list of elements as positional arguments
(e.g., \textsl{{-}{-}include\_interstitials Mn}).
To obtain both intrinsic and extrinsic interstitials
the intrinsic elements have to be explicitly mentioned
(e.g., \textsl{{-}{-}include\_interstitials Ga As Mn}).
As for the other defect types,
PyCDT enumerates the interstitial calculation folders
according to symmetrically distinct sites found.
The \textsl{info} part of the interstitial folder names
indicate (1) the type of the atom
located on an interstitial site having a certain 
(2) coordination pattern and (3) chemical environment.
For example,
\textsl{inter\ul{\ \ }1\ul{\ \ }As\ul{\ \ }oct\ul{\ \ }Ga6}
shows that we are dealing with an As interstitial that is
octahedrally coordinated by six Ga atoms.

For each defect type in semiconductors, multiple charge states are considered according to 
the algorithm outlined in Sec.~\ref{chargeassignmentsection}. For insulators
 a conservative charge assignment
is used as described in Sec.~\ref{chargeassignmentsection}. By default, the input structure
is considered a semiconductor. To specify the input structure is of insulator type, the
option 
\textsl{{-}{-}type
insulator} can be used.
The user can also
modify the charge assignments for each defect by specifying 
either of the two flags, \textsl{{-}{-}oxi\ul{\ \ }state} or 
\textsl{{-}{-}oxi\ul{\ \ }range}.
Alternatively, the option \textsl{{-}{-}type manual allows for the user to specify every charge
state that is desired.}
The DFT input files associated with each of these charge states,
\textsl{q}, are deposited into subfolders named \textsl{charge\ul{  }q}.
For example, seven charge states are generated for the gallium vacancy in GaAs.

Apart from the structure file,
PyCDT automatically generates all other input files
according to the settings used for the MP. The input 
settings can also be easily modified by 
supplying the parameters in a yaml or json file and using the keyword
\textsl{-{}-input\ul{\ \ }settings\ul{\ \ }file $\langle$settings\ul{\ \ }file.yaml$\rangle$}. 
For instance, the file \textsl{user\ul{\ \ }settings.yaml} that we provide in
the \textsl{examples} folder
changes the default functional from PBE to PBEsol and increases
the energy cutoff to 620~eV. 
When such changes are made, the user has to keep in 
mind that the atomic chemical potentials obtained from the MP database in the final parsing step 
have to be replaced by 
user computed ones with the corresponding changes included during chemical
potential calculations
 ~- a process which can be sped up substantially with PyCDT's phase diagram set up and parsing feature.
Any DFT settings that are specific to 
either the bulk, or the dielectric, or the defect calculations
can be thus realized, too, as demonstrated in the example file.
The input settings whether chosen by default or by the user are expected to be
tested for appropriate convergence criteria. 
In addition to the input files for DFT calculations, PyCDT 
saves a \textsl{transformation.json} in each calculation folder,
except for \textsl{dielectric}, to facilitate post-processing.

\subsection{Parsing Finished Calculations}

The DFT calculations can be run either manually,
with bash scripts, or with high throughput frameworks~\cite{fireworks}.
Once all the calculations have successfully completed, the generated output files 
are parsed to obtain all the data needed to compute defect formation energies, $E^\mathrm{f}$. 
This part of PyCDT is executed by reading the \emph{transformation.json} file that was output from 
the previous file generation step. 
To initiate parsing from the command line interface, the user issues:
\begin{framed}
\begin{table}[H]
\begin{tabular}{lllll}
    $\triangleright$pycdt parse\underline{ }output & [ &{-}{-}directory & $\langle$directory$\rangle$ &] \\
                                    & [ &-{}-mpid & $\langle$mpid$\rangle$&]\\
                                    & [ &-{}-mapi\ul{\ \ }key &$\langle$mapi\ul{\ \ }key$\rangle$&] \\
\end{tabular}
\end{table}
\end{framed}
Here, directory is the root folder of the calculations. If executed within the folder of the calculations the option can be omitted. 
Once the parsing is completed, PyCDT stores all the data required for next steps in a file called ``defect\ul{\ \ }data.json''.
If any of the calculations were not successfully converged according to the code output files,
PyCDT raises a warning, but continues parsing the rest of the calculations.
The output file ``defect\ul{\ \ }data.json" contains the parsed energies of 
the defect and bulk supercells as well as other information required in the 
next steps to calculate finite size charge corrections and defect formation 
energies. Some of the additional data such as the dielectric constant is 
obtained by parsing the output from the dielectric calculation. 
The band gap and atomic chemical potentials generated in this step are obtained
from computed entries in the MP database. Note that these band gaps 
are only accurate at the level of GGA, which often underestimates the gap by about 50\%. The output file ``defect\ul{\ \ }data.json" 
is highly readable and users can edit the file to supplant parameters either parsed from the DFT calculations or obtained from the MP database.
If the user prefers to have formation energies and transition levels closer to a higher 
level of theory (which can provide better band gaps than the GGA approximation), the band edge 
alignment procedures suggested in
Section~\ref{supercellsection} can be used.
Furthermore note
that some structures in the MP database
do not have
fully computed band structures,
which results
in poorly converged band-gap characteristics.

\subsection{Computation of Correction Term}

\begin{figure}[!ht]
  \begin{center}
    \includegraphics[width=7.8cm]{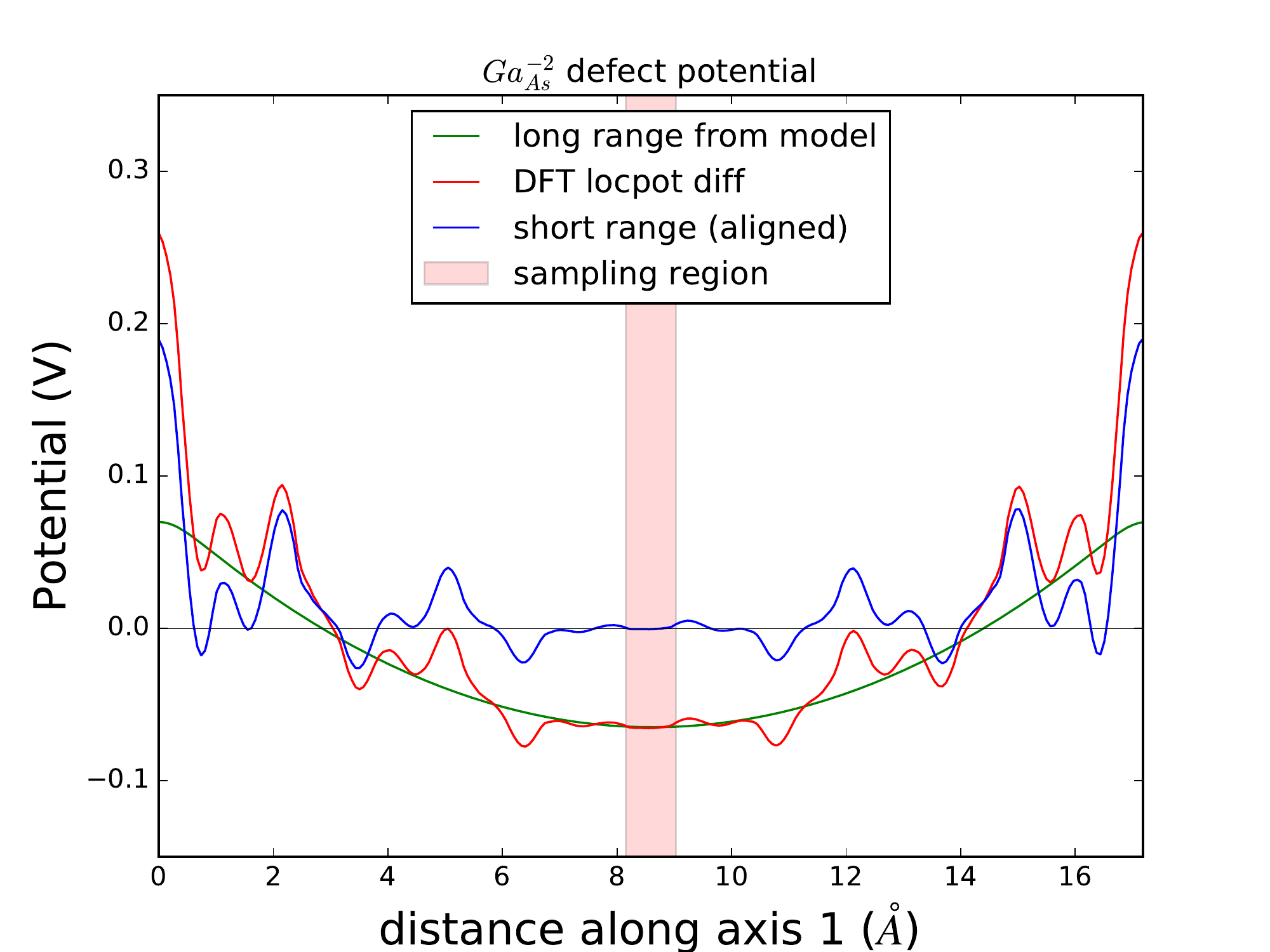}
    \includegraphics[width=7.8cm]{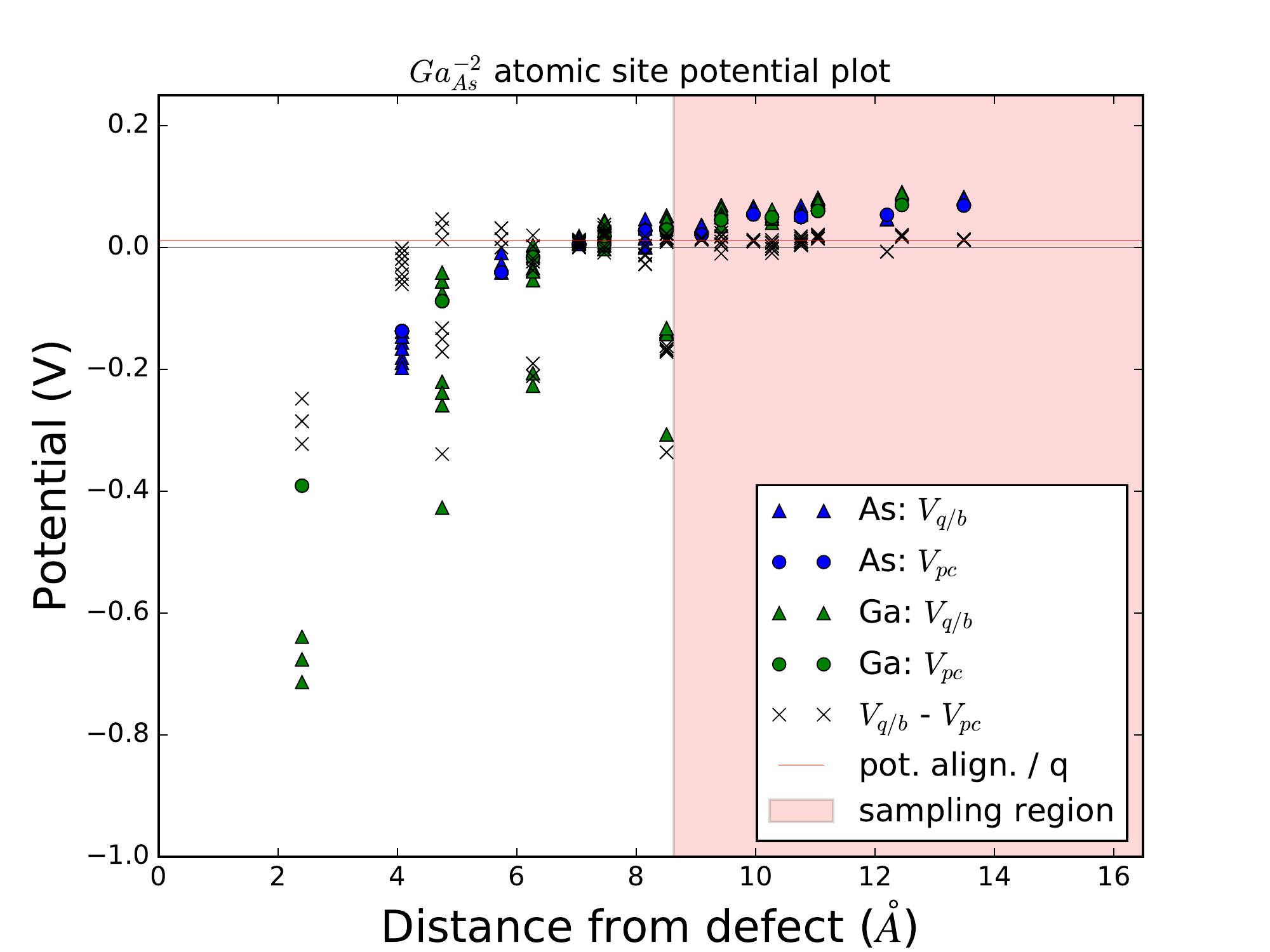}
    \caption{\label{fig:GaAscorrectionplots}Two different methods for computing the potential alignment correction on GaAs calculation. At left is isotropic correction, developed by Freysoldt \emph{et al.}, using the planar average method~\cite{freysoldt2009,freysoldt2011}, and at right is anisotropic correction by Kumagai and Oba, using the atomic site averaging method.~\cite{kumagai2014}. }
  \end{center}
\end{figure}

Correcting the errors due to long-range Coulomb interactions in finite-size supercells 
results in improved defect formation energies.  A feature of PyCDT is the
possibility to compute such corrections with minimal work from the end user. 
The following command-line call computes individual correction values
for all charged defects found in the present directory:\\
\begin{framed}
\begin{table}[H]
\begin{tabular}{lllll}
 $\triangleright$pycdt compute\underline{ }corrections & [ &{-}{-}correction\underline{ }method & $\langle$correction method$\rangle$&] \\
                     & [ &{-}{-}epsilon& $\langle$epsilon tensor$\rangle$&] \\
                     & [ &{-}{-}input\underline{ }file\underline{ }name & $\langle$defect data file$\rangle$&] \\
                     & [ &{-}{-}plot\underline{ }results& ] \\
\end{tabular}
\end{table}
\end{framed}
Here, correction method keywords can be either \textsl{freysoldt} for 
correction due to Freysoldt \emph{et al.}~\cite{freysoldt2009,freysoldt2011} or \textsl{kumagai} 
for the approach extended to anisotropic 
systems by Kumagai and Oba~\cite{kumagai2014}. 
As shown in Section~\ref{validCorrs}, these codes have been rigorously 
tested against the results from the codes of the original authors.
The command line interface requires the defect data file generated in the previous step, \textsl{defect\ul{\ \ }data.json}, for computing the corrections. 
This flag can be omitted if the file name is unchanged. 
The calculated corrections for each defect charge state are stored in a file called \textsl{corrections.json}.
By rerunning the command with different correction keywords, one can quickly obtain the corrections computed with different 
frameworks. Shown in Figure \ref{fig:GaAscorrectionplots} are the resulting potential alignment plots for 
each correction type on the $Ga_{As}^{-2}$ defect in GaAs. 

The sampling regions for obtaining the potential alignment correction defaults to 1$\AA$ in 
the middle region of the planar average plots 
recommended by Freysoldt \emph{et al.}~\cite{freysoldt2011}, 
and to the region outside of 
the Wigner-Seitz radius for atomic site averaging method, following the approach described in Ref.~\cite{kumagai2014}.
The width of these default sampling regions can be changed by modifying 
the instantiation of the relevant PyCDT correction classes.

\subsection{Formation Energy Plots and Transition Levels}

\begin{figure}[!ht]
  \begin{center}
    \includegraphics[width=7.8cm]{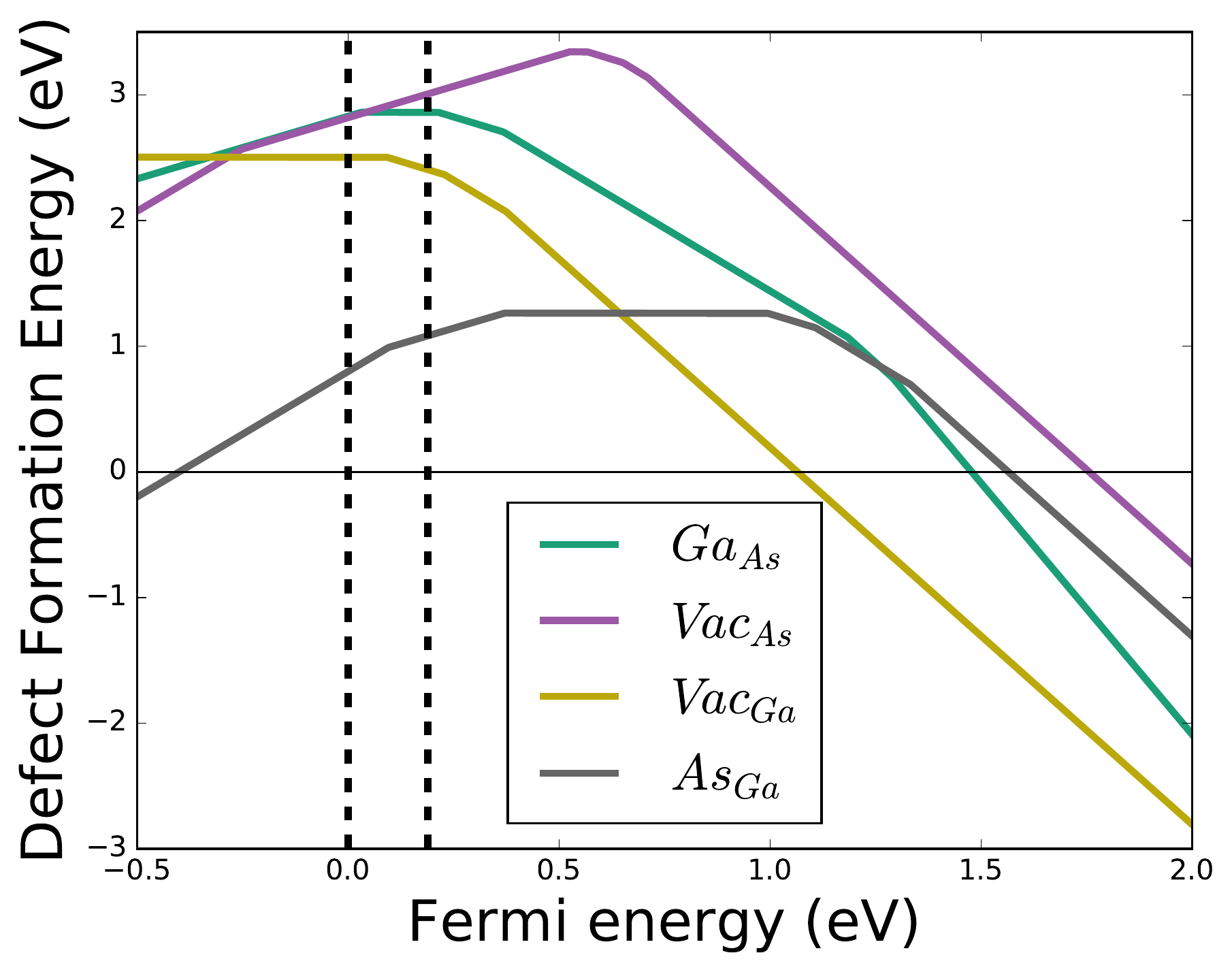}
    \includegraphics[width=7.8cm]{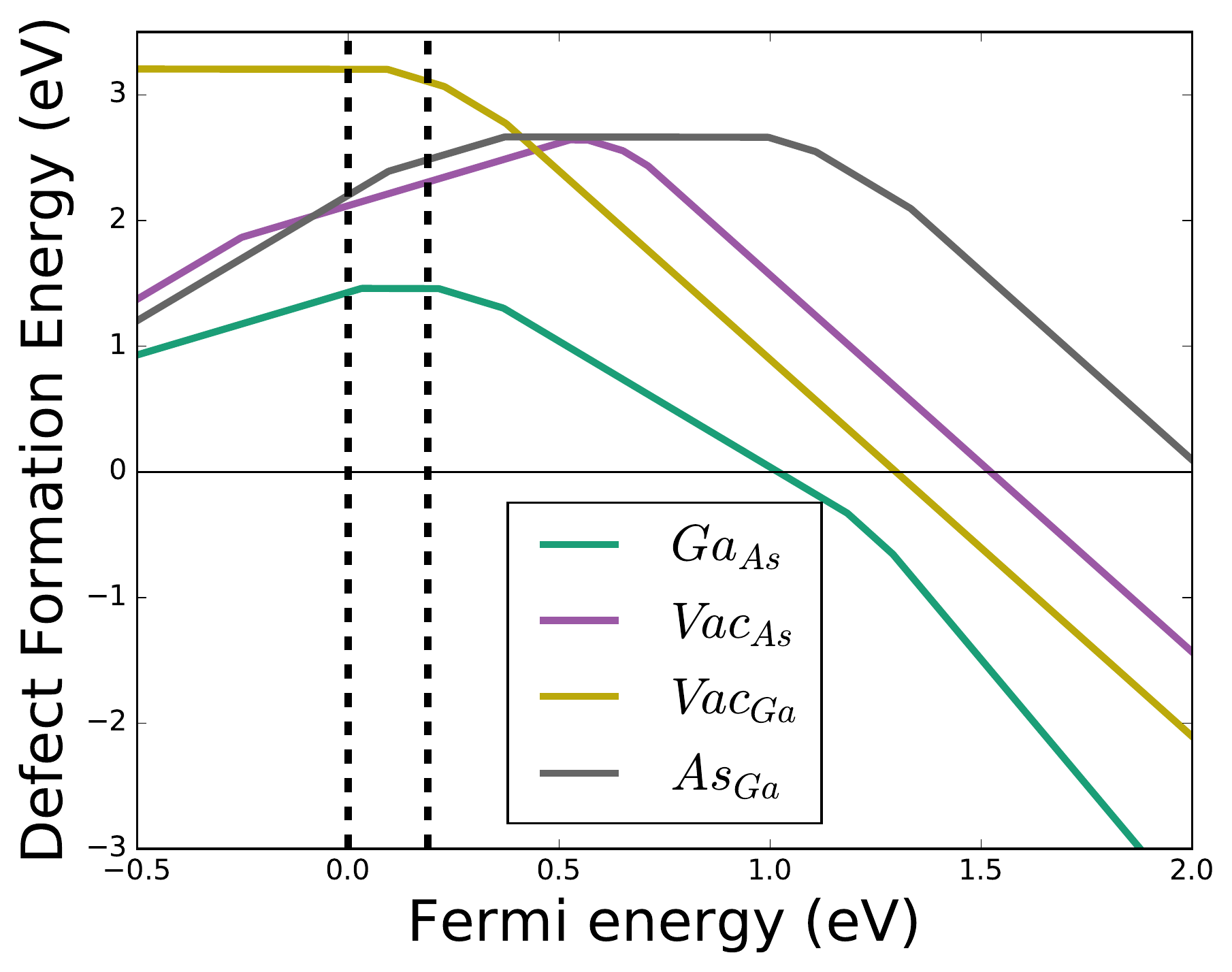}
    \caption{\label{fig:GaAsformenplots} Defect formation-energy plots from PyCDT for GaAs.
        The left panel is obtained in the As-rich growth regime,
        whereas the right panel is obtained under Ga-rich conditions.
        The dotted black line indicates the GGA-PBE band gap of
        GaAs~\cite{jain2013commentary}.}
  \end{center}
\end{figure}

Once the defect energetics and the correction values are obtained
and specified in the \textsl{defect\ul{\ \ }data.json} and
\textsl{corrections.json} files, the transitition levels and
formation energies of each defect across the band gap
can be determined by:
\begin{framed}
\begin{table}[H]
\begin{tabular}{lllll}
$\triangleright$pycdt compute\underline{ }formation\underline{ }energies & [ &{-}{-}input\underline{ }file\underline{ }name & $\langle$defectsdata json file$\rangle$&]\\
							       & [ &{-}{-}corrections\underline{ }file\underline{ }name & $\langle$corrections json file$\rangle$&] \\
							       & [ &{-}{-}bandgap & $\langle$band gap$\rangle$&] \\
                     & [ &{-}{-}plot\underline{ }results& ] \\
\end{tabular}
\end{table}
\end{framed}
By default, PyCDT uses the band gap stored in the MP database,
which is computed with GGA-PBE and, hence, under-predicted when 
compared to the experimental band gap. 
As mentioned several times in this work, this approximation can be improved upon with the optional 
corrections included, such as the band edge realignment feature of PyCDT.
A simpler approximation to improving the defect formation energies is
to plot the defect formation energies 
across the experimental gap the user can specify the experimental band gap. This can be done from the command line
with the keyword \textsl{--bandgap $\langle$band gap$\rangle$}. 
This has the effect of extending the gap by shifting the
conduction band minimum,
but keeping the position of the valence band maximum
and the defect levels fixed,
and it is often called the ``extended gap'' scheme.
We note that the extended gap option is strictly for plotting purposes and, thus, does not 
alter the defect formation energies nor the transition levels.  If the names 
of the files obtained in the previous two steps are not changed, the 
corresponding options can be omitted.  If the \textsl{corrections.json} file is not found 
and an alternative corrections file is not specified, PyCDT assumes that 
electrostatic corrections are not desired and computes the defect formation 
energies without any corrections.  As detailed in Section~\ref{chemicalpotentialsection},
the defect formation energies are 
influenced by the chemical environment, and their range is determined by 
the phase stability of the various compounds formed by the constituent host 
and defect elements.  Hence, two plots are generated corresponding to the 
chemical availability of the constituent host elements in the compound. The 
files "Ga\ul{\ \ }rich\ul{\ \ }formation\ul{\ \ }energy.eps" and 
"As\ul{\ \ }rich\ul{\ \ }formation\ul{\ \ }energy.eps" are shown in 
Figure~\ref{fig:GaAsformenplots}. 
These results are verified to be consistent with literature results in Section~\ref{verifGaAs}.

\section{Validation and Verification}\label{sec::valid}
\subsection{Validation}\label{validCorrs}
To validate PyCDT's implementation of each correction method, we ran the correction 
methods on the defects generated and computed for 21 zinc blende structures (binary 
and elemental), comparing the charge corrections generated by PyCDT with the 
open-source code developed by Freysoldt \emph{et al.}, \textsc{sxdefectalign}, as 
well as with the command line code developed by Kumagai and Oba. Over a total of 
192 defects calculated, the root mean square difference (absolute average relative errors) between PyCDT and the 
original author's codes were 6.4 meV (1.5\%) and 14.4 meV (3.4\%) for the correction by Freysoldt 
\emph{et al.} and the correction by Kumagai and Oba, respectively.
The differences 
in the corrections are almost entirely attributed to the resolution in the 
calculation of the potential correction. 

\subsection{Verification}\label{verifGaAs}
To verify the results predicted from the example GaAs test set,
we compared the defect formation energies in GaAs obtained from PyCDT with the
data reported in literature.  We note that all of the transition levels for 
the $Vac_{Ga}$ and $As_{Ga}$ defects are within the range of transition levels 
that we found for semi-local functional approximations in the literature.
An exception is the $As_{Ga}$(-1/-2) transition,
which appears very far into the GGA-PBE conduction band.
This outlier is off by 0.273~eV relative to the reported results 
by Chroneos \emph{et al}~\cite{chroneos}.
For all transition levels that we 
predicted, we find a root mean square deviation of 0.218 eV from the window of 
values found in the literature~\cite{tahini, chroneos, schultz, El-Mellouhi, Schick, 
KomsaMicroelec, northrup1994GaAs, komsa2011GaAs}. This is a modest 
variation that reflects the difficulty in predicting defect levels
consistently---even within the same level of theory.

\section{Conclusion}

We have introduced PyCDT,
a Python toolkit which facilitates the setup
and post-processing of point defect calculations
of semiconductor and insulator materials
with widely available DFT suites.
This open source code allows for coupling automated defect calculations 
to the massive amount of data generated by the Materials Project database.
Apart from the underlying theory, approaches, and algorithms,
this paper presents a detailed guide for how to use PyCDT at every step
of the computation of charged defect properties
employing the well-studied example of GaAs. 
While the example results were obtained from VASP~\cite{kresse1993ab,kresse1994ab}
calculations,
we carefully developed PyCDT in an
abstracted form that adopts the advantageous
code agnosticism of pymatgen~\cite{ong2013python}.
This makes the provided tools attractive to any user
interested in running defect calculations, 
regardless of DFT code preference.
However, we emphasize that, despite its convenience and
our effort to construct sensible defaults,
the computation of defect properties with PyCDT still requires user expertise
for choosing appropriate settings in certain circumstances and
for interpreting the results meaningfully in general
(i.e., we discourage purely ``black box'' usage).

The PyCDT version presented here is 1.0.0.
Future updates will, amongst others, include adaptations related to
improved charge delocalization analysis,
further defect corrections for issues like
artificial band dispersion, as well as the 
possibility of generating defect complexes.
On the application side, PyCDT could also be extended to compute configuration 
coordinate diagrams, so as to evaluate optical and luminescence transitions associated 
with the point defects in materials targeting optical applications.

We hope that our openly available tools will help to standardize computational research in the
realm of charged defects.
In particular, we hope that reproducibility issues
commonly encountered in DFT calculations~\cite{lejaeghere2016}
can be more effectively identified and tackled.

\section{Acknowledgments}
This work was intellectually led by the Materials Project Center, supported by 
the Office of Basic Energy Sciences (BES) of the U.S. Department of Energy 
(DOE) under Grant No. EDCBEE. 
 B. Medasani was supported by the U.S. DOE, Office of BES, Division of 
Materials Sciences and Engineering. Pacific Northwest National Laboratory is 
a multiprogram national laboratory operated for DOE by Battelle. This work 
used resources of the National Energy Research Scientific Computing Center, 
supported by the BES of the U.S. DOE under Contract No. DE-AC02-05CH11231. 
The authors would like to thank 
Yu Kumagai for insightful discussions about charge corrections.

\appendix

\newpage
\section{Charge States from Literature}
\label{sec:charge:lit}

\begin{center}
\begin{longtable}{c c c c c}
\caption{\label{tab:charge:states} Charge states from literature;
brackets refer to hybrid functional rather
than (semi-)local functional results.} \\
\hline
Structure  &        Defect         & \multicolumn{2}{c}{Charge States}  &  Ref.  \\
           &                       &    from          &       to           &        \\
\hline
\endfirsthead
\hline 
Structure  &        Defect         & \multicolumn{2}{c}{Charge States}  &  Ref.  \\
           &                       &    from          &       to           &        \\
\hline
\endhead
\hline
\multicolumn{5}{r}{\emph{To be continued on next page.}} \\
\endfoot
\hline
\endlastfoot
C          &   V$_\mathrm{C}$    &  $+2$  &  $-2$  &  \citenum{deak}     \\
           &   N$_\mathrm{C}$    &  $+1$  &  $-1$  &                     \\
AlP        &   V$_\mathrm{Al}$   &  $0$   &  $-3$  &  \citenum{tahini}   \\
           &   V$_\mathrm{P}$    &  $+1$  &  $-2$  &                     \\
           &   Al$_\mathrm{P}$   &  $+1$  &  $-2$  &  \citenum{chroneos} \\
           &   P$_\mathrm{Al}$   &  $+2$  &  $-2$  &                     \\
AlAs       &   V$_\mathrm{Al}$   &  $0$   &  $-3$  &  \citenum{tahini}   \\
           &   V$_\mathrm{As}$   &  $+1$  &  $-2$  &                     \\
           &   Al$_\mathrm{As}$  &  $+1$  &  $-2$  &  \citenum{chroneos} \\
           &   As$_\mathrm{Al}$  &  $+1$  &  $-1$  &                     \\
AlSb       &   V$_\mathrm{Al}$   &  $0$   &  $-3$  &  \citenum{tahini}   \\
           &   V$_\mathrm{Sb}$   &  $+1$  &  $-3$  &                     \\
           &   Al$_\mathrm{Sb}$  &  $0$   &  $-2$  &  \citenum{chroneos} \\
           &   Sb$_\mathrm{Al}$  &  $+1$  &  $-1$  &                     \\
Si         &   V$_\mathrm{Si}$   &  $+2$  &  $-2$  &  \citenum{corsetti} \\
ZnS        &   V$_\mathrm{S}$    &  $+2$  &  $0$   &  \citenum{li2012}   \\ 
           &   V$_\mathrm{Zn}$   &  $0$ ($+1$)   &  $-2$  &  \citenum{li2012} (\citenum{petretto2015})   \\
           &   S$_\mathrm{Zn}$   &  $0$   &  $-2$  &  \citenum{li2012}   \\
           &   Zn$_\mathrm{S}$   &  $+2$  &  $0$   &                     \\
           &   Zn$_i$            &  $+2$  &  $0$   &                     \\
           &    S$_i$            &  $0$   &  $-2$  &                     \\
ZnSe       &   V$_\mathrm{Se}$   &  $+2$  &  $-2$  &  \citenum{dossantos2011}   \\
           &   V$_\mathrm{Zn}$   &  $+2$  &  $-2$  &                     \\
           &   Cl$_\mathrm{Se}$  &  $+2$  &  $-1$  &                     \\
           &   F$_\mathrm{Se}$   &  $+1$  &  $-2$  &                     \\
           &   F$_\mathrm{Zn}$   &  $-2$  &  $-2$  &                     \\
ZnTe       &   V$_\mathrm{Zn}$   &  $0$ ($+1$)   &  $-2$  &  \citenum{petretto2015}  \\
GaN        &   V$_\mathrm{Ga}$   &  $0$ ($+1$)   &  $-3$  &  \citenum{neugebauer1994} (\citenum{petretto2015}) \\
           &   V$_\mathrm{N}$    &  $+1$  &  $+1$  &  \citenum{neugebauer1994} \\
           &   Ga$_\mathrm{N}$   &  $+1$  &  $-2$  &                     \\
           &   N$_\mathrm{Ga}$   &  $+2$  &  $-1$  &                     \\
           &   N$_i$             &  $+3$  &  $-1$  &                     \\
           &   Ga$_i$            &  $+3$  &  $+1$  &                     \\
           &   C$_\mathrm{N}$    &  $0$ ($+1$)   &  $-1$  & \citenum{petretto2015} \\
GaP        &   V$_\mathrm{Ga}$   &  $0$   &  $-3$  &  \citenum{tahini}   \\
           &   V$_\mathrm{P}$    &  $+1$  &  $-3$  &                     \\
           &   Ga$_\mathrm{P}$   &  $0$   &  $-2$  &  \citenum{chroneos} \\
           &   P$_\mathrm{Ga}$   &  $+2$  &  $-2$  &                     \\ 
GaAs       &   V$_\mathrm{Ga}$   &  $-1$  &  $-3$  &  \citenum{tahini}   \\
           &   V$_\mathrm{As}$   &  $+1$  &  $-3$  &                     \\
           &   Ga$_\mathrm{As}$  &  $0$   &  $-3$  &  \citenum{chroneos} \\
           &   As$_\mathrm{Ga}$  &  $+1$  &  $-2$  &                     \\
GaSb       &   V$_\mathrm{Ga}$   &  $0$   &  $-3$  &  \citenum{tahini}   \\
           &   V$_\mathrm{Sb}$   &  $0$   &  $-3$  &                     \\
           &   Ga$_\mathrm{Sb}$  &  $0$   &  $-2$  &  \citenum{chroneos} \\
           &   Sb$_\mathrm{Ga}$  &  $+1$  &  $-1$  &                     \\
CdS        &   V$_\mathrm{S}$    &  $+2$  &  $0$   &  \citenum{wu2011}   \\
           &   V$_\mathrm{Cd}$   &  $0$   &  $-2$  &                     \\
           &   Cd$_\mathrm{S}$   &  $+2$  &  $+2$  &                     \\
           &   S$_\mathrm{Cd}$   &  $+4$  &  $-2$  &                     \\
           &   Cd$_i$            &  $+2$  &  $+2$  &                     \\
           &   S$_i$             &  $+4$  &  $-2$  &                     \\
           &   Mn$_\mathrm{Cd}$  &  $+1$  &  $0$   &                     \\
           &   Fe$_\mathrm{Cd}$  &  $+2$  &  $0$   &                     \\
           &   Co$_\mathrm{Cd}$  &  $+1$  &  $0$   &                     \\
           &   Ni$_\mathrm{Cd}$  &  $+1$  &  $0$   &                     \\
           &   Mn$_i$            &  $+3$  &  $+2$  &                     \\
           &   Fe$_i$            &  $+3$  &  $+2$  &                     \\
           &   Co$_i$            &  $+3$  &  $+2$  &                     \\
           &   Ni$_i$            &  $+2$  &  $+1$  &                     \\
InP        &   V$_\mathrm{In}$   &  $0$   &  $-3$  &  \citenum{tahini}   \\
           &   V$_\mathrm{P}$    &  $+1$  &  $-1$  &                     \\
           &   In$_\mathrm{P}$   &  $+1$  &  $-2$  &  \citenum{chroneos} \\
           &   P$_\mathrm{In}$   &  $+2$  &  $-1$  &                     \\
InAs       &   V$_\mathrm{In}$   &  $0$   &  $-2$  &  \citenum{tahini}   \\
           &   V$_\mathrm{As}$   &  $+1$  &  $0$   &                     \\
           &   In$_\mathrm{As}$  &  $0$   &  $-1$  &  \citenum{chroneos} \\
           &   As$_\mathrm{In}$  &  $+1$  &  $0$   &                     \\
InSb       &   V$_\mathrm{In}$   &  $-1$  &  $-2$  &  \citenum{tahini}   \\
           &   V$_\mathrm{Sb}$   &  $+1$  &  $0$   &                     \\
           &   In$_\mathrm{Sb}$  &  $0$   &  $-1$  &  \citenum{chroneos} \\
           &   Sb$_\mathrm{In}$  &  $+1$  &  $0$   &                     \\
\end{longtable}
\end{center}

\section{List of Acronyms}
\label{sec:acr}

\begin{longtable}{r l}
\hline \bfseries Acronym & \bfseries Full Form \\ \hline
\endfirsthead
\hline \bfseries Acronym & \bfseries Full Form \\ \hline
\endhead
\hline
\multicolumn{2}{r}{\emph{To be continued on next page.}}
\endfoot
\hline
\endlastfoot
DFT    &    electronic density functional theory  \\
GGA    &    generalized gradient approximation \\
LDA    &    local-density approximation \\
MAPI    &    Materials Application Programming Interface  \\
MP    &    The Materials Project \\  
MPID    &    Materials Project identifier \\
PBC    &    periodic boundary conditions  \\
PBE    &    Perdew--Burke--Ernzerhof  \\
PyCDT    &    Python Charged Defect Toolkit  \\
Pymatgen    &    Python Materials Genomics  \\
VASP    &    Vienna Ab initio Simulation Package  \\
\end{longtable}


\section{Reference}
\bibliographystyle{elsarticle-num}
\bibliography{reference}

\begin{thebibliography}{10}
\expandafter\ifx\csname url\endcsname\relax
  \def\url#1{\texttt{#1}}\fi
\expandafter\ifx\csname urlprefix\endcsname\relax\def\urlprefix{URL }\fi
\expandafter\ifx\csname href\endcsname\relax
  \def\href#1#2{#2} \def\path#1{#1}\fi

\bibitem{numpy}
Numpy developers, http://numpy.org/.

\bibitem{matplotlib}
J.~D. Hunter, Matplotlib: A 2d graphics environment, Computing In Science \&
  Engineering 9~(3) (2007) 90--95.

\bibitem{ong2013python}
S.~P. Ong, W.~D. Richards, A.~Jain, G.~Hautier, M.~Kocher, S.~Cholia,
  D.~Gunter, V.~L. Chevrier, K.~A. Persson, G.~Ceder, Python materials genomics
  (pymatgen): A robust, open-source python library for materials analysis,
  Computational Materials Science 68 (2013) 314--319.

\bibitem{callister2007}
W.~D. Callister, Jr., {Materials Science and Engineering: an Introduction}, 7th
  Edition, John Wiley \& Sons, Inc., New York, NY, USA, 2007.

\bibitem{queisser2013}
H.~Queisser, J.~Spaeth, H.~Overhof, {Point Defects in Semiconductors and
  Insulators: Determination of Atomic and Electronic Structure from
  Paramagnetic Hyperfine Interactions}, Springer Series in Materials Science,
  Springer Berlin Heidelberg, 2013.

\bibitem{mccluskey2012}
M.~McCluskey, E.~Haller, {Dopants and Defects in Semiconductors}, CRC Press,
  2012.

\bibitem{rodnyi1997}
P.~Rodnyi, {Physical Processes in Inorganic Scintillators}, Laser \& Optical
  Science \& Technology, Taylor \& Francis, 1997.

\bibitem{seebauer2006}
E.~G. Seebauer, M.~C. Kratzer, Charged point defects in semiconductors, Mater.
  Sci. Eng., R 55~(3-6) (2006) 57--149.
\newblock \href {http://dx.doi.org/{10.1016/j.mser.2006.01.002}}
  {\path{doi:{10.1016/j.mser.2006.01.002}}}.

\bibitem{freysoldt_rmp}
C.~Freysoldt, B.~Grabowski, T.~Hickel, J.~Neugebauer, G.~Kresse, A.~Janotti,
  C.~G. Van~de Walle,
  \href{http://link.aps.org/doi/10.1103/RevModPhys.86.253}{First-principles
  calculations for point defects in solids}, Rev. Mod. Phys. 86 (2014)
  253--305.
\newblock \href {http://dx.doi.org/10.1103/RevModPhys.86.253}
  {\path{doi:10.1103/RevModPhys.86.253}}.
\newline\urlprefix\url{http://link.aps.org/doi/10.1103/RevModPhys.86.253}

\bibitem{dorenbos2005}
P.~Dorenbos, \href{http://dx.doi.org/10.1002/pssa.200460106}{Scintillation
  mechanisms in {Ce}$^{3+}$ doped halide scintillators}, physica status solidi
  (a) 202~(2) (2005) 195--200.
\newblock \href {http://dx.doi.org/10.1002/pssa.200460106}
  {\path{doi:10.1002/pssa.200460106}}.
\newline\urlprefix\url{http://dx.doi.org/10.1002/pssa.200460106}

\bibitem{chaudhry2014}
A.~Chaudhry, R.~Boutchko, S.~Chourou, G.~Zhang, N.~Gr\o{}nbech-Jensen,
  A.~Canning,
  \href{http://link.aps.org/doi/10.1103/PhysRevB.89.155105}{First-principles
  study of luminescence in {Eu}${}^{2+}$-doped inorganic scintillators}, Phys.
  Rev. B 89 (2014) 155105.
\newblock \href {http://dx.doi.org/10.1103/PhysRevB.89.155105}
  {\path{doi:10.1103/PhysRevB.89.155105}}.
\newline\urlprefix\url{http://link.aps.org/doi/10.1103/PhysRevB.89.155105}

\bibitem{zunger_TCO}
S.~Lany, A.~Zunger,
  \href{http://link.aps.org/doi/10.1103/PhysRevLett.98.045501}{{Dopability,
  Intrinsic Conductivity, and Nonstoichiometry of Transparent Conducting
  Oxides}}, Phys. Rev. Lett. 98 (2007) 045501.
\newblock \href {http://dx.doi.org/10.1103/PhysRevLett.98.045501}
  {\path{doi:10.1103/PhysRevLett.98.045501}}.
\newline\urlprefix\url{http://link.aps.org/doi/10.1103/PhysRevLett.98.045501}

\bibitem{scanlon_TCO}
D.~O. Scanlon, G.~W. Watson,
  \href{http://dx.doi.org/10.1021/jz1011725}{{Conductivity Limits in CuAlO$_2$
  from Screened-Hybrid Density Functional Theory}}, The Journal of Physical
  Chemistry Letters 1~(21) (2010) 3195--3199.
\newblock \href {http://arxiv.org/abs/http://dx.doi.org/10.1021/jz1011725}
  {\path{arXiv:http://dx.doi.org/10.1021/jz1011725}}, \href
  {http://dx.doi.org/10.1021/jz1011725} {\path{doi:10.1021/jz1011725}}.
\newline\urlprefix\url{http://dx.doi.org/10.1021/jz1011725}

\bibitem{varley_TCO}
J.~B. Varley, V.~Lordi, A.~Miglio, G.~Hautier,
  \href{http://link.aps.org/doi/10.1103/PhysRevB.90.045205}{{Electronic
  structure and defect properties of B${}_{6}$O from hybrid functional and
  many-body perturbation theory calculations: A possible ambipolar transparent
  conductor}}, Phys. Rev. B 90 (2014) 045205.
\newblock \href {http://dx.doi.org/10.1103/PhysRevB.90.045205}
  {\path{doi:10.1103/PhysRevB.90.045205}}.
\newline\urlprefix\url{http://link.aps.org/doi/10.1103/PhysRevB.90.045205}

\bibitem{zakutayev_solar}
A.~Zakutayev, C.~M. Caskey, A.~N. Fioretti, D.~S. Ginley, J.~Vidal,
  V.~Stevanovic, E.~Tea, S.~Lany,
  \href{http://dx.doi.org/10.1021/jz5001787}{{Defect Tolerant Semiconductors
  for Solar Energy Conversion}}, The Journal of Physical Chemistry Letters
  5~(7) (2014) 1117--1125, pMID: 26274458.
\newblock \href {http://arxiv.org/abs/http://dx.doi.org/10.1021/jz5001787}
  {\path{arXiv:http://dx.doi.org/10.1021/jz5001787}}, \href
  {http://dx.doi.org/10.1021/jz5001787} {\path{doi:10.1021/jz5001787}}.
\newline\urlprefix\url{http://dx.doi.org/10.1021/jz5001787}

\bibitem{walsh_solar}
A.~Walsh, D.~O. Scanlon, S.~Chen, X.~G. Gong, S.-H. Wei,
  \href{http://dx.doi.org/10.1002/ange.201409740}{Self-regulation mechanism for
  charged point defects in hybrid halide perovskites}, Angewandte Chemie
  127~(6) (2015) 1811--1814.
\newblock \href {http://dx.doi.org/10.1002/ange.201409740}
  {\path{doi:10.1002/ange.201409740}}.
\newline\urlprefix\url{http://dx.doi.org/10.1002/ange.201409740}

\bibitem{zhu_te}
H.~Zhu, G.~Hautier, U.~Aydemir, Z.~M. Gibbs, G.~Li, S.~Bajaj, J.-H. Pohls,
  D.~Broberg, W.~Chen, A.~Jain, M.~A. White, M.~Asta, G.~J. Snyder, K.~Persson,
  G.~Ceder, \href{http://dx.doi.org/10.1039/C5TC01440A}{{Computational and
  experimental investigation of TmAgTe${}_{2}$ and XYZ$_{2}$ compounds{,} a new
  group of thermoelectric materials identified by first-principles
  high-throughput screening}}, J. Mater. Chem. C 3 (2015) 10554--10565.
\newblock \href {http://dx.doi.org/10.1039/C5TC01440A}
  {\path{doi:10.1039/C5TC01440A}}.
\newline\urlprefix\url{http://dx.doi.org/10.1039/C5TC01440A}

\bibitem{pomrehn_te}
G.~S. Pomrehn, A.~Zevalkink, W.~G. Zeier, A.~van de Walle, G.~J. Snyder,
  \href{http://dx.doi.org/10.1002/anie.201311125}{{Defect-Controlled Electronic
  Properties in {A}{Z}n$_{2}${S}b${}_{2}$ Zintl Phases}}, Angewandte Chemie
  International Edition 53~(13) (2014) 3422--3426.
\newblock \href {http://dx.doi.org/10.1002/anie.201311125}
  {\path{doi:10.1002/anie.201311125}}.
\newline\urlprefix\url{http://dx.doi.org/10.1002/anie.201311125}

\bibitem{vandewalle_jap}
C.~G. Van~de Walle, J.~Neugebauer,
  \href{http://scitation.aip.org/content/aip/journal/jap/95/8/10.1063/1.1682673}{{First-principles
  calculations for defects and impurities: Applications to {III}-nitrides}},
  Journal of Applied Physics 95~(8) (2004) 3851--3879.
\newblock \href {http://dx.doi.org/http://dx.doi.org/10.1063/1.1682673}
  {\path{doi:http://dx.doi.org/10.1063/1.1682673}}.
\newline\urlprefix\url{http://scitation.aip.org/content/aip/journal/jap/95/8/10.1063/1.1682673}

\bibitem{zunger}
S.~Lany, A.~Zunger,
  \href{http://stacks.iop.org/0965-0393/17/i=8/a=084002}{Accurate prediction of
  defect properties in density functional supercell calculations}, Modelling
  and Simulation in Materials Science and Engineering 17~(8) (2009) 084002.
\newline\urlprefix\url{http://stacks.iop.org/0965-0393/17/i=8/a=084002}

\bibitem{peng_gap}
H.~Peng, D.~O. Scanlon, V.~Stevanovic, J.~Vidal, G.~W. Watson, S.~Lany,
  Convergence of density and hybrid functional defect calculations for compound
  semiconductors, Physical Review B 88~(11) (2013) 115201.

\bibitem{bruneval_GW}
M.~Giantomassi, M.~Stankovski, R.~Shaltaf, M.~Gr{\"u}ning, F.~Bruneval,
  P.~Rinke, G.-M. Rignanese, Electronic properties of interfaces and defects
  from many-body perturbation theory: Recent developments and applications,
  physica status solidi (b) 248~(2) (2011) 275--289.

\bibitem{northrup_GW}
J.~E. Northrup, M.~S. Hybertsen, S.~G. Louie,
  \href{http://link.aps.org/doi/10.1103/PhysRevLett.59.819}{Theory of
  quasiparticle energies in alkali metals}, Phys. Rev. Lett. 59 (1987)
  819--822.
\newblock \href {http://dx.doi.org/10.1103/PhysRevLett.59.819}
  {\path{doi:10.1103/PhysRevLett.59.819}}.
\newline\urlprefix\url{http://link.aps.org/doi/10.1103/PhysRevLett.59.819}

\bibitem{li_GW}
J.-L. Li, G.-M. Rignanese, E.~K. Chang, X.~Blase, S.~G. Louie,
  \href{http://link.aps.org/doi/10.1103/PhysRevB.66.035102}{$\mathrm{GW}$ study
  of the metal-insulator transition of bcc hydrogen}, Phys. Rev. B 66 (2002)
  035102.
\newblock \href {http://dx.doi.org/10.1103/PhysRevB.66.035102}
  {\path{doi:10.1103/PhysRevB.66.035102}}.
\newline\urlprefix\url{http://link.aps.org/doi/10.1103/PhysRevB.66.035102}

\bibitem{ldau}
A.~I. Liechtenstein, V.~I. Anisimov, J.~Zaanen,
  \href{https://link.aps.org/doi/10.1103/PhysRevB.52.R5467}{Density-functional
  theory and strong interactions: Orbital ordering in mott-hubbard insulators},
  Phys. Rev. B 52 (1995) R5467--R5470.
\newblock \href {http://dx.doi.org/10.1103/PhysRevB.52.R5467}
  {\path{doi:10.1103/PhysRevB.52.R5467}}.
\newline\urlprefix\url{https://link.aps.org/doi/10.1103/PhysRevB.52.R5467}

\bibitem{TiO2_defects}
E.~Finazzi, C.~Di~Valentin, G.~Pacchioni, A.~Selloni, {Excess electron states
  in reduced bulk anatase TiO$_2$: comparison of standard GGA, GGA+U, and
  hybrid DFT calculations.}, The Journal of {C}hemical {P}hysics 129~(15)
  (2008) 154113--154113.

\bibitem{wu2011}
J.-C. Wu, J.~Zheng, P.~Wu, R.~Xu, Study of native defects and transition-metal
  ({Mn}, {Fe}, {Co}, and {Ni}) doping in a zinc-blende {CdS} photocatalyst by
  {DFT} and hybrid {DFT} calculations, J. Phys. Chem. C 115~(13) (2011)
  5675--5682.
\newblock \href {http://dx.doi.org/{10.1021/jp109567c}}
  {\path{doi:{10.1021/jp109567c}}}.

\bibitem{castleton_review}
C.~W.~M. Castleton, A.~Höglund, S.~Mirbt,
  \href{http://stacks.iop.org/0965-0393/17/i=8/a=084003}{Density functional
  theory calculations of defect energies using supercells}, Modelling and
  Simulation in Materials Science and Engineering 17~(8) (2009) 084003.
\newline\urlprefix\url{http://stacks.iop.org/0965-0393/17/i=8/a=084003}

\bibitem{goyal2016computational}
A.~Goyal, P.~Gorai, H.~Peng, S.~Lany, V.~Stevanovi{\'c}, A computational
  framework for automation of point defect calculations, Computational
  Materials Science 130 (2017) 1--9.

\bibitem{pean2017presentation}
E.~P{\'e}an, J.~Vidal, S.~Jobic, C.~Latouche, Presentation of the {PyDEF}
  post-treatment python software to compute publishable charts for defect
  energy formation, Chemical Physics Letters.

\bibitem{yim2017property}
K.~Yim, J.~Lee, D.~Lee, M.~Lee, E.~Cho, H.~Lee, H.~Nahm, S.~Han, Property
  database for single-element doping in {ZnO} obtained by automated
  first-principles calculations., Scientific {R}eports 7 (2017) 40907--40907.

\bibitem{jain2013commentary}
A.~Jain, S.~P. Ong, G.~Hautier, W.~Chen, W.~D. Richards, S.~Dacek, S.~Cholia,
  D.~Gunter, D.~Skinner, G.~Ceder, et~al., Commentary: The materials project: A
  materials genome approach to accelerating materials innovation, APL Materials
  1~(1) (2013) 011002.

\bibitem{pbe_functional}
J.~P. Perdew, K.~Burke, M.~Ernzerhof, Generalized gradient approximation made
  simple, Physical review letters 77~(18) (1996) 3865.

\bibitem{freysoldt2009}
C.~Freysoldt, J.~Neugebauer, C.~G. Van~de Walle,
  \href{http://link.aps.org/doi/10.1103/PhysRevLett.102.016402}{{Fully
  \textit{Ab Initio} Finite-Size Corrections for Charged-Defect Supercell
  Calculations}}, Phys. Rev. Lett. 102 (2009) 016402.
\newblock \href {http://dx.doi.org/10.1103/PhysRevLett.102.016402}
  {\path{doi:10.1103/PhysRevLett.102.016402}}.
\newline\urlprefix\url{http://link.aps.org/doi/10.1103/PhysRevLett.102.016402}

\bibitem{kumagai2014}
Y.~Kumagai, F.~Oba,
  \href{http://link.aps.org/doi/10.1103/PhysRevB.89.195205}{Electrostatics-based
  finite-size corrections for first-principles point defect calculations},
  Phys. Rev. B 89 (2014) 195205.
\newblock \href {http://dx.doi.org/10.1103/PhysRevB.89.195205}
  {\path{doi:10.1103/PhysRevB.89.195205}}.
\newline\urlprefix\url{http://link.aps.org/doi/10.1103/PhysRevB.89.195205}

\bibitem{zimmermann2017}
N.~E.~R. Zimmermann, A.~Jain, M.~Haranczyk, Structure motif assessment based on
  order parameters for automatic motif identification, interstitial finding,
  and diffusion path characterization, in preparation (2017).

\bibitem{kresse1993ab}
G.~Kresse, J.~Hafner, Ab initio molecular dynamics for liquid metals, Physical
  Review B 47~(1) (1993) 558.

\bibitem{kresse1994ab}
G.~Kresse, J.~Hafner, Ab initio molecular-dynamics simulation of the
  liquid-metal--amorphous-semiconductor transition in germanium, Physical
  Review B 49~(20) (1994) 14251.

\bibitem{ss_electrolytes}
W.~D. Richards, L.~J. Miara, Y.~Wang, J.~C. Kim, G.~Ceder, Interface stability
  in solid-state batteries, Chemistry of Materials 28~(1) (2015) 266--273.

\bibitem{metaGGASCAN}
J.~Sun, R.~C. Remsing, Y.~Zhang, Z.~Sun, A.~Ruzsinszky, H.~Peng, Z.~Yang,
  A.~Paul, U.~Waghmare, X.~Wu, et~al., Accurate first-principles structures and
  energies of diversely bonded systems from an efficient density functional,
  Nature {C}hemistry 8~(9) (2016) 831--836.

\bibitem{hybridreview}
B.~G. Janesko, T.~M. Henderson, G.~E. Scuseria,
  \href{http://dx.doi.org/10.1039/B812838C}{Screened hybrid density functionals
  for solid-state chemistry and physics}, Phys. Chem. Chem. Phys. 11 (2009)
  443--454.
\newblock \href {http://dx.doi.org/10.1039/B812838C}
  {\path{doi:10.1039/B812838C}}.
\newline\urlprefix\url{http://dx.doi.org/10.1039/B812838C}

\bibitem{tahini}
H.~A. Tahini, A.~Chroneos, S.~T. Murphy, U.~Schwingenschl\"{o}gl, R.~W. Grimes,
  Vacancies and defect levels in {III}-{V} semiconductors, J. Appl. Phys.
  114~(6).
\newblock \href {http://dx.doi.org/{10.1063/1.4818484}}
  {\path{doi:{10.1063/1.4818484}}}.

\bibitem{medvedevscience2017}
M.~G. Medvedev, I.~S. Bushmarinov, J.~Sun, J.~P. Perdew, K.~A. Lyssenko,
  \href{http://science.sciencemag.org/content/355/6320/49}{Density functional
  theory is straying from the path toward the exact functional}, Science
  355~(6320) (2017) 49--52.
\newblock \href
  {http://arxiv.org/abs/http://science.sciencemag.org/content/355/6320/49.full.pdf}
  {\path{arXiv:http://science.sciencemag.org/content/355/6320/49.full.pdf}},
  \href {http://dx.doi.org/10.1126/science.aah5975}
  {\path{doi:10.1126/science.aah5975}}.
\newline\urlprefix\url{http://science.sciencemag.org/content/355/6320/49}

\bibitem{wolvertonNainPbTe}
J.~W. Doak, K.~J. Michel, C.~Wolverton, {Determining dilute-limit solvus
  boundaries in multi-component systems using defect energetics: Na in PbTe and
  PbS}, Journal of Materials Chemistry C 3~(40) (2015) 10630--10649.

\bibitem{zhang}
S.~B. Zhang, J.~E. Northrup,
  \href{http://link.aps.org/doi/10.1103/PhysRevLett.67.2339}{{Chemical
  potential dependence of defect formation energies in GaAs: Application to Ga
  self-diffusion}}, Phys. Rev. Lett. 67 (1991) 2339--2342.
\newblock \href {http://dx.doi.org/10.1103/PhysRevLett.67.2339}
  {\path{doi:10.1103/PhysRevLett.67.2339}}.
\newline\urlprefix\url{http://link.aps.org/doi/10.1103/PhysRevLett.67.2339}

\bibitem{jain2011b}
A.~Jain, G.~Hautier, S.~P. Ong, C.~J. Moore, C.~C. Fischer, K.~A. Persson,
  G.~Ceder,
  \href{https://link.aps.org/doi/10.1103/PhysRevB.84.045115}{Formation
  enthalpies by mixing {GGA} and {GGA $+$ $U$} calculations}, Phys. Rev. B 84
  (2011) 045115.
\newblock \href {http://dx.doi.org/10.1103/PhysRevB.84.045115}
  {\path{doi:10.1103/PhysRevB.84.045115}}.
\newline\urlprefix\url{https://link.aps.org/doi/10.1103/PhysRevB.84.045115}

\bibitem{medasani2017}
B.~Medasani, M.~L. Sushko, K.~M. Rosso, D.~K. Schreiber, S.~M. Bruemmer,
  \href{http://dx.doi.org/10.1021/acs.jpcc.7b00071}{Vacancies and
  vacancy-mediated self diffusion in {Cr$_2$O$_3$}: A first-principles study},
  The Journal of Physical Chemistry C 121~(3) (2017) 1817--1831.
\newblock \href
  {http://arxiv.org/abs/http://dx.doi.org/10.1021/acs.jpcc.7b00071}
  {\path{arXiv:http://dx.doi.org/10.1021/acs.jpcc.7b00071}}, \href
  {http://dx.doi.org/10.1021/acs.jpcc.7b00071}
  {\path{doi:10.1021/acs.jpcc.7b00071}}.
\newline\urlprefix\url{http://dx.doi.org/10.1021/acs.jpcc.7b00071}

\bibitem{tahini2016}
H.~A. Tahini, X.~Tan, U.~Schwingenschlögl, S.~C. Smith,
  \href{http://dx.doi.org/10.1021/acscatal.6b00937}{Formation and migration of
  oxygen vacancies in {SrCoO$_3$} and their effect on oxygen evolution
  reactions}, ACS Catalysis 6~(8) (2016) 5565--5570.
\newblock \href
  {http://arxiv.org/abs/http://dx.doi.org/10.1021/acscatal.6b00937}
  {\path{arXiv:http://dx.doi.org/10.1021/acscatal.6b00937}}, \href
  {http://dx.doi.org/10.1021/acscatal.6b00937}
  {\path{doi:10.1021/acscatal.6b00937}}.
\newline\urlprefix\url{http://dx.doi.org/10.1021/acscatal.6b00937}

\bibitem{agoston_vibentropy}
P.~{\'A}goston, K.~Albe, R.~M. Nieminen, M.~J. Puska, {Intrinsic n-type
  behavior in transparent conducting oxides: A comparative hybrid-functional
  study of In${}_2$O${}_3$, SnO${}_2$, and ZnO}, Physical review letters
  103~(24) (2009) 245501.

\bibitem{hautier_dftenergies}
G.~Hautier, S.~P. Ong, A.~Jain, C.~J. Moore, G.~Ceder,
  \href{http://link.aps.org/doi/10.1103/PhysRevB.85.155208}{Accuracy of density
  functional theory in predicting formation energies of ternary oxides from
  binary oxides and its implication on phase stability}, Phys. Rev. B 85 (2012)
  155208.
\newblock \href {http://dx.doi.org/10.1103/PhysRevB.85.155208}
  {\path{doi:10.1103/PhysRevB.85.155208}}.
\newline\urlprefix\url{http://link.aps.org/doi/10.1103/PhysRevB.85.155208}

\bibitem{metastable_gerd}
W.~Sun, S.~T. Dacek, S.~P. Ong, G.~Hautier, A.~Jain, W.~D. Richards, A.~C.
  Gamst, K.~A. Persson, G.~Ceder,
  \href{http://advances.sciencemag.org/content/2/11/e1600225}{The thermodynamic
  scale of inorganic crystalline metastability}, Science Advances 2~(11).
\newblock \href
  {http://arxiv.org/abs/http://advances.sciencemag.org/content/2/11/e1600225.full.pdf}
  {\path{arXiv:http://advances.sciencemag.org/content/2/11/e1600225.full.pdf}},
  \href {http://dx.doi.org/10.1126/sciadv.1600225}
  {\path{doi:10.1126/sciadv.1600225}}.
\newline\urlprefix\url{http://advances.sciencemag.org/content/2/11/e1600225}

\bibitem{lany_fs}
S.~Lany, A.~Zunger, {Assessment of correction methods for the band-gap problem
  and for finite-size effects in supercell defect calculations: Case studies
  for ZnO and GaAs}, Physical Review B 78~(23) (2008) 235104.

\bibitem{taylor_fs}
S.~E. Taylor, F.~Bruneval, Understanding and correcting the spurious
  interactions in charged supercells, Physical Review B 84~(7) (2011) 075155.

\bibitem{freysoldt2011}
C.~Freysoldt, J.~Neugebauer, C.~G. Van~de Walle,
  \href{http://dx.doi.org/10.1002/pssb.201046289}{Electrostatic interactions
  between charged defects in supercells}, physica status solidi (b) 248~(5)
  (2011) 1067--1076.
\newblock \href {http://dx.doi.org/10.1002/pssb.201046289}
  {\path{doi:10.1002/pssb.201046289}}.
\newline\urlprefix\url{http://dx.doi.org/10.1002/pssb.201046289}

\bibitem{MakovPayne_correction}
G.~Makov, M.~C. Payne,
  \href{http://link.aps.org/doi/10.1103/PhysRevB.51.4014}{Periodic boundary
  conditions in \textit{ab initio} calculations}, Phys. Rev. B 51 (1995)
  4014--4022.
\newblock \href {http://dx.doi.org/10.1103/PhysRevB.51.4014}
  {\path{doi:10.1103/PhysRevB.51.4014}}.
\newline\urlprefix\url{http://link.aps.org/doi/10.1103/PhysRevB.51.4014}

\bibitem{komsa2012}
H.-P. Komsa, T.~T. Rantala, A.~Pasquarello,
  \href{http://link.aps.org/doi/10.1103/PhysRevB.86.045112}{Finite-size
  supercell correction schemes for charged defect calculations}, Phys. Rev. B
  86 (2012) 045112.
\newblock \href {http://dx.doi.org/10.1103/PhysRevB.86.045112}
  {\path{doi:10.1103/PhysRevB.86.045112}}.
\newline\urlprefix\url{http://link.aps.org/doi/10.1103/PhysRevB.86.045112}

\bibitem{boeck2011}
S.~Boeck, C.~Freysoldt, A.~Dick, L.~Ismer, J.~Neugebauer,
  \href{http://www.sciencedirect.com/science/article/pii/S0010465510003619}{{The
  object-oriented DFT program library S/PHI/nX}}, Computer Physics
  Communications 182~(3) (2011) 543 -- 554.
\newblock \href {http://dx.doi.org/http://dx.doi.org/10.1016/j.cpc.2010.09.016}
  {\path{doi:http://dx.doi.org/10.1016/j.cpc.2010.09.016}}.
\newline\urlprefix\url{http://www.sciencedirect.com/science/article/pii/S0010465510003619}

\bibitem{petretto2015}
G.~Petretto, F.~Bruneval, Systematic defect donor levels in {III}-{V} and
  {II}-{VI} semiconductors revealed by hybrid functional density-functional
  theory, Phys. Rev. B 92~(22) (2015) 224111.
\newblock \href {http://dx.doi.org/{10.1103/PhysRevB.92.224111}}
  {\path{doi:{10.1103/PhysRevB.92.224111}}}.

\bibitem{deak}
P.~De\'{a}k, B.~Aradi, M.~Kaviani, T.~Frauenheim, A.~Gali, Formation of {NV}
  centers in diamond: a theoretical study based on calculated transitions and
  migration of nitrogen and vacancy related defects, Phys. Rev. B 89~(7) (2014)
  075203.
\newblock \href {http://dx.doi.org/{10.1103/PhysRevB.89.075203}}
  {\path{doi:{10.1103/PhysRevB.89.075203}}}.

\bibitem{chroneos}
A.~Chroneos, H.~A. Tahini, U.~Schwingenschl\"{o}gl, R.~W. Grimes, Antisites in
  {III}-{V} semiconductors: {D}ensity functional theory calculations, J. Appl.
  Phys. 116~(2).
\newblock \href {http://dx.doi.org/{10.1063/1.4887135}}
  {\path{doi:{10.1063/1.4887135}}}.

\bibitem{corsetti}
F.~Corsetti, A.~A. Mostofi, System-size convergence of point defect properties:
  the case of the silicon vacancy, Phys. Rev. B 84~(3) (2011) 035209.
\newblock \href {http://dx.doi.org/{10.1103/PhysRevB.84.035209}}
  {\path{doi:{10.1103/PhysRevB.84.035209}}}.

\bibitem{li2012}
P.~Li, S.~Deng, L.~Zhang, G.~Liu, J.~Yu, Native point defects in {ZnS}:
  first-principles studies based on {LDA}, {LDA} $+$ {U} and an extrapolation
  scheme, Chem. Phys. Lett. 531 (2012) 75--79.
\newblock \href {http://dx.doi.org/{10.1016/j.cplett.2012.02.008}}
  {\path{doi:{10.1016/j.cplett.2012.02.008}}}.

\bibitem{dossantos2011}
L.~S. dos Santos, W.~G. Schmidt, E.~Rauls, Group-{VII} point defects in {ZnSe},
  Phys. Rev. B 84~(11) (2011) 115201.
\newblock \href {http://dx.doi.org/{10.1103/PhysRevB.84.115201}}
  {\path{doi:{10.1103/PhysRevB.84.115201}}}.

\bibitem{neugebauer1994}
J.~Neugebauer, C.~G. Van~de Walle, Atomic geometry and electronic structure of
  native defects in {GaN}, Phys. Rev. B 50~(11) (1994) 8067--8070.
\newblock \href {http://dx.doi.org/{10.1103/PhysRevB.50.8067}}
  {\path{doi:{10.1103/PhysRevB.50.8067}}}.

\bibitem{okeeffe1991}
M.~O'Keeffe, N.~E. Brese, Atom sizes and bond lengths in molecules and
  crystals, J. Am. Chem. Soc. 113~(9) (1991) 3226--3229.
\newblock \href {http://dx.doi.org/{10.1021/ja00009a002}}
  {\path{doi:{10.1021/ja00009a002}}}.

\bibitem{gibbson1994}
A.~Gibson, R.~Haydock, J.~P. LaFemina,
  \href{http://link.aps.org/doi/10.1103/PhysRevB.50.2582}{{Stability of vacancy
  defects in MgO: The role of charge neutrality}}, Phys. Rev. B 50 (1994)
  2582--2592.
\newblock \href {http://dx.doi.org/10.1103/PhysRevB.50.2582}
  {\path{doi:10.1103/PhysRevB.50.2582}}.
\newline\urlprefix\url{http://link.aps.org/doi/10.1103/PhysRevB.50.2582}

\bibitem{peters2009}
B.~Peters, {Competing nucleation pathways in a mixture of oppositely charged
  colloids: out-of-equilibrium nucleation revisited}, J. Chem. Phys. 131~(24)
  (2009) 244103.

\bibitem{zimmermann2015}
N.~E.~R. Zimmermann, B.~Vorselaars, D.~Quigley, B.~Peters, Nucleation of {NaCl}
  from aqueous solution: critical sizes, ion-attachment kinetics, and rates, J.
  Am. Chem. Soc. 137~(41) (2015) 13352--13361.

\bibitem{Decoster:ApplPhysLett:2008}
S.~Decoster, B.~De~Vries, U.~Wahl, J.~G. Correia, A.~Vantomme, Experimental
  evidence of tetrahedral interstitial and bond-centered {Er} in {Ge}, Appl.
  Phys. Lett. 93~(14) (2008) 141907.
\newblock \href {http://dx.doi.org/{10.1063/1.2996280}}
  {\path{doi:{10.1063/1.2996280}}}.

\bibitem{Decoster:PhysRevLett:2009}
S.~Decoster, S.~Cottenier, B.~De~Vries, H.~Emmerich, U.~Wahl, J.~G. Correia,
  A.~Vantomme, Transition metal impurities on the bond-centered site in
  germanium, Phys. Rev. Lett. 102~(6) (2009) 065502.
\newblock \href {http://dx.doi.org/{10.1103/PhysRevLett.102.065502}}
  {\path{doi:{10.1103/PhysRevLett.102.065502}}}.

\bibitem{Decoster:JApplPhys:2009}
S.~Decoster, B.~De~Vries, U.~Wahl, J.~G. Correia, A.~Vantomme, Lattice location
  study of implanted {In} in {Ge}, J. Appl. Phys. 105~(8) (2009) 083522.
\newblock \href {http://dx.doi.org/{10.1063/1.3110104}}
  {\path{doi:{10.1063/1.3110104}}}.

\bibitem{Decoster:PhysRevB:2010}
S.~Decoster, S.~Cottenier, U.~Wahl, J.~G. Correia, A.~Vantomme, Lattice
  location study of ion implanted {Sn} and {Sn}-related defects in {Ge}, Phys.
  Rev. B 81~(15) (2010) 155204.
\newblock \href {http://dx.doi.org/{10.1103/PhysRevB.81.155204}}
  {\path{doi:{10.1103/PhysRevB.81.155204}}}.

\bibitem{Decoster:ApplPhysLett:2010}
S.~Decoster, S.~Cottenier, U.~Wahl, J.~G. Correia, L.~M.~C. Pereira,
  C.~Lacasta, M.~R. Da~Silva, A.~Vantomme, Diluted manganese on the
  bond-centered site in germanium, Appl. Phys. Lett. 97~(15) (2010) 151914.
\newblock \href {http://dx.doi.org/{10.1063/1.3501123}}
  {\path{doi:{10.1063/1.3501123}}}.

\bibitem{Pereira:ApplPhysLett:2011}
L.~M.~C. Pereira, U.~Wahl, S.~Decoster, J.~G. Correia, M.~R. da~Silva,
  A.~Vantomme, J.~P. Ara\'{u}jo, Direct identification of interstitial {Mn} in
  heavily $p$-type doped {GaAs} and evidence of its high thermal stability,
  Appl. Phys. Lett. 98~(20) (2011) 201905.
\newblock \href {http://dx.doi.org/{10.1063/1.3592568}}
  {\path{doi:{10.1063/1.3592568}}}.

\bibitem{Pereira:PhysRevB:2012}
L.~M.~C. Pereira, U.~Wahl, S.~Decoster, J.~G. Correia, L.~M. Amorim, M.~R.
  da~Silva, J.~P. Araujo, A.~Vantomme, Stability and diffusion of interstitial
  and substitutional {Mn} in {GaAs} of different doping types, Phys. Rev. B
  86~(12) (2012) 125206.
\newblock \href {http://dx.doi.org/{10.1103/PhysRevB.86.125206}}
  {\path{doi:{10.1103/PhysRevB.86.125206}}}.

\bibitem{Decoster:JApplPhys:2012}
S.~Decoster, U.~Wahl, S.~Cottenier, J.~G. Correia, T.~Mendon\c{c}a, L.~M.
  Amorim, L.~M.~C. Pereira, A.~Vantomme, Lattice position and thermal stability
  of diluted {As} in {Ge}, J. Appl. Phys. 111~(5) (2012) 053528.
\newblock \href {http://dx.doi.org/{10.1063/1.3692761}}
  {\path{doi:{10.1063/1.3692761}}}.

\bibitem{Amorim:ApplPhysLett:2013}
L.~M. Amorim, U.~Wahl, L.~M.~C. Pereira, S.~Decoster, D.~J. Silva, M.~R.
  da~Silva, A.~Gottberg, J.~G. Correia, K.~Temst, A.~Vantomme, Precise lattice
  location of substitutional and interstitial {Mg} in {AlN}, Appl. Phys. Lett.
  103~(26) (2013) 262102.
\newblock \href {http://dx.doi.org/{10.1063/1.4858389}}
  {\path{doi:{10.1063/1.4858389}}}.

\bibitem{Silva:JApplPhys:2014}
D.~J. Silva, U.~Wahl, J.~G. Correia, L.~M.~C. Pereira, L.~M. Amorim, M.~R.
  da~Silva, E.~Bosne, J.~P. Ara\'{u}jo, Lattice location and thermal stability
  of implanted nickel in silicon studied by on-line emission channeling, J.
  Appl. Phys. 115~(2) (2014) 023504.
\newblock \href {http://dx.doi.org/{10.1063/1.4861142}}
  {\path{doi:{10.1063/1.4861142}}}.

\bibitem{Hofsass:PhysRepRevSecPhysLett:1991}
H.~Hofs\"{a}ss, G.~Lindner, Emission channeling and blocking, Phys. Rep. (Rev.
  Sec. Phys. Lett.) 201~(3) (1991) 121--183.
\newblock \href {http://dx.doi.org/{10.1016/0370-1573(91)90121-2}}
  {\path{doi:{10.1016/0370-1573(91)90121-2}}}.

\bibitem{Silva:RevSciInstrum:2013}
M.~R. Silva, U.~Wahl, J.~G. Correia, L.~M. Amorim, L.~M.~C. Pereira, A
  versatile apparatus for on-line emission channeling experiments, Rev. Sci.
  Instrum. 84~(7) (2013) 073506.
\newblock \href {http://dx.doi.org/{10.1063/1.4813266}}
  {\path{doi:{10.1063/1.4813266}}}.

\bibitem{Rong:ChemMater:2015}
Z.~Rong, R.~Malik, P.~Canepa, G.~S. Gautam, M.~Liu, A.~Jain, K.~Persson,
  G.~Ceder, Materials design rules for multivalent ion mobility in
  intercalation structures, Chem. Mater. 27~(17) (2015) 6016--6021.
\newblock \href {http://dx.doi.org/{10.1021/acs.chemmater.5b02342}}
  {\path{doi:{10.1021/acs.chemmater.5b02342}}}.

\bibitem{Wang:NatMater:2015}
Y.~Wang, W.~D. Richards, S.~P. Ong, L.~J. Miara, J.~C. Kim, Y.~Mo, G.~Ceder,
  Design principles for solid-state lithium superionic conductors, Nat. Mater.
  14~(10) (2015) 1026--1031.
\newblock \href {http://dx.doi.org/{10.1038/NMAT4369}}
  {\path{doi:{10.1038/NMAT4369}}}.

\bibitem{devita1992}
A.~De~Vita, The energetics of defects and impurities in metals and ionic
  materials from first principles, Ph.D. thesis, Keele University (Aug 1992).

\bibitem{fireworks}
A.~Jain, S.~P. Ong, W.~Chen, B.~Medasani, X.~Qu, M.~Kocher, M.~Brafman,
  G.~Petretto, G.-M. Rignanese, G.~Hautier, D.~Gunter, K.~A. Persson,
  \href{http://dx.doi.org/10.1002/cpe.3505}{Fireworks: a dynamic workflow
  system designed for high-throughput applications}, Concurrency and
  Computation: Practice and Experience 27~(17) (2015) 5037--5059,
  cPE-14-0307.R2.
\newblock \href {http://dx.doi.org/10.1002/cpe.3505}
  {\path{doi:10.1002/cpe.3505}}.
\newline\urlprefix\url{http://dx.doi.org/10.1002/cpe.3505}

\bibitem{schultz}
P.~A. Schultz, O.~A. von Lilienfeld,
  \href{http://stacks.iop.org/0965-0393/17/i=8/a=084007}{Simple intrinsic
  defects in gallium arsenide}, Modelling and Simulation in Materials Science
  and Engineering 17~(8) (2009) 084007.
\newline\urlprefix\url{http://stacks.iop.org/0965-0393/17/i=8/a=084007}

\bibitem{El-Mellouhi}
F.~El-Mellouhi, N.~Mousseau,
  \href{http://link.aps.org/doi/10.1103/PhysRevB.71.125207}{{Self-vacancies in
  gallium arsenide: An \textit{ab initio} calculation}}, Phys. Rev. B 71 (2005)
  125207.
\newblock \href {http://dx.doi.org/10.1103/PhysRevB.71.125207}
  {\path{doi:10.1103/PhysRevB.71.125207}}.
\newline\urlprefix\url{http://link.aps.org/doi/10.1103/PhysRevB.71.125207}

\bibitem{Schick}
J.~T. Schick, C.~G. Morgan, P.~Papoulias,
  \href{http://link.aps.org/doi/10.1103/PhysRevB.66.195302}{{First-principles
  study of As interstitials in GaAs: Convergence, relaxation, and formation
  energy}}, Phys. Rev. B 66 (2002) 195302.
\newblock \href {http://dx.doi.org/10.1103/PhysRevB.66.195302}
  {\path{doi:10.1103/PhysRevB.66.195302}}.
\newline\urlprefix\url{http://link.aps.org/doi/10.1103/PhysRevB.66.195302}

\bibitem{KomsaMicroelec}
H.-P. Komsa, A.~Pasquarello,
  \href{http://www.sciencedirect.com/science/article/pii/S0167931711003406}{{Identification
  of defect levels at As/oxide interfaces through hybrid functionals}},
  Microelectronic Engineering 88~(7) (2011) 1436 -- 1439, proceedings of the
  17th Biennial International Insulating Films on Semiconductor Conference17th
  Biennial International Insulating Films on Semiconductor Conference.
\newblock \href {http://dx.doi.org/http://dx.doi.org/10.1016/j.mee.2011.03.081}
  {\path{doi:http://dx.doi.org/10.1016/j.mee.2011.03.081}}.
\newline\urlprefix\url{http://www.sciencedirect.com/science/article/pii/S0167931711003406}

\bibitem{northrup1994GaAs}
J.~E. Northrup, S.~B. Zhang,
  \href{http://link.aps.org/doi/10.1103/PhysRevB.50.4962}{{Energetics of the As
  vacancy in GaAs: The stability of the 3+ charge state}}, Phys. Rev. B 50
  (1994) 4962--4964.
\newblock \href {http://dx.doi.org/10.1103/PhysRevB.50.4962}
  {\path{doi:10.1103/PhysRevB.50.4962}}.
\newline\urlprefix\url{http://link.aps.org/doi/10.1103/PhysRevB.50.4962}

\bibitem{komsa2011GaAs}
H.-P. Komsa, A.~Pasquarello,
  \href{http://link.aps.org/doi/10.1103/PhysRevB.84.075207}{{Assessing the
  accuracy of hybrid functionals in the determination of defect levels:
  Application to the As antisite in GaAs}}, Phys. Rev. B 84 (2011) 075207.
\newblock \href {http://dx.doi.org/10.1103/PhysRevB.84.075207}
  {\path{doi:10.1103/PhysRevB.84.075207}}.
\newline\urlprefix\url{http://link.aps.org/doi/10.1103/PhysRevB.84.075207}

\bibitem{lejaeghere2016}
K.~Lejaeghere, G.~Bihlmayer, T.~Bjoerkman, P.~Blaha, S.~Bluegel, V.~Blum,
  D.~Caliste, I.~E. Castelli, S.~J. Clark, A.~Dal~Corso, S.~de~Gironcoli,
  T.~Deutsch, J.~K. Dewhurst, I.~Di~Marco, C.~Draxl, M.~Dulak, O.~Eriksson,
  J.~A. Flores-Livas, K.~F. Garrity, L.~Genovese, P.~Giannozzi, M.~Giantomassi,
  S.~Goedecker, X.~Gonze, O.~Granaes, E.~K.~U. Gross, A.~Gulans, F.~Gygi, D.~R.
  Hamann, P.~J. Hasnip, N.~A.~W. Holzwarth, D.~Iusan, D.~B. Jochym, F.~Jollet,
  D.~Jones, G.~Kresse, K.~Koepernik, E.~Kuecuekbenli, Y.~O. Kvashnin, I.~L.~M.
  Locht, S.~Lubeck, M.~Marsman, N.~Marzari, U.~Nitzsche, L.~Nordstrom,
  T.~Ozaki, L.~Paulatto, C.~J. Pickard, W.~Poelmans, M.~I.~J. Probert,
  K.~Refson, M.~Richter, G.-M. Rignanese, S.~Saha, M.~Scheffler, M.~Schlipf,
  K.~Schwarz, S.~Sharma, F.~Tavazza, P.~Thunstroem, A.~Tkatchenko, M.~Torrent,
  D.~Vanderbilt, M.~J. van Setten, V.~Van~Speybroeck, J.~M. Wills, J.~R. Yates,
  G.-X. Zhang, S.~Cottenier, Reproducibility in density functional theory
  calculations of solids, Science 351~(6280) (2016) 1415--U81.

\end{thebibliography}







\end{document}